\documentclass[journal]{IEEEtran}

\usepackage{amsmath,epsfig}
\usepackage[T1]{fontenc}
\usepackage{psfrag}
\usepackage{array}
\usepackage{cite}
\usepackage{amsfonts}
\usepackage{algorithm}
\usepackage{algorithmic}
\usepackage[utf8]{inputenc} 

\usepackage{amsmath}
\usepackage{color}
\usepackage{fancybox}
\usepackage{array}
\usepackage[dvipsnames]{xcolor}
\definecolor{MyGreen}{rgb}{0.4660 0.6740 0.1880}
\definecolor{MyPink}{rgb}{1 0.6 0.7373}

\usepackage{graphicx}

\usepackage[caption=false,listofformat=subsimple,labelformat=simple]{subfig}

\newcolumntype{P}[1]{>{\centering\arraybackslash}p{#1}}
\newcolumntype{M}[1]{>{\centering\arraybackslash}m{#1}}

\usepackage{circuitikz}
\usepackage{tikz}
\usetikzlibrary{fit,positioning}
\usetikzlibrary{shapes.geometric}
\usepackage{adjustbox}

\usepackage[justification=centering]{caption}
\usepackage{float}
\usepackage{url}

\DeclareFontFamily{OT1}{pzc}{}
\DeclareFontShape{OT1}{pzc}{m}{it}%
             {<-> s * [1.28] pzcmi7t}{} 
\DeclareMathAlphabet{\freqDom}{OT1}{pzc}
                                 {m}{it}
\usepackage{amsbsy} 

\usepackage{tcolorbox}
\tcbuselibrary{listingsutf8} 

\usepackage{soul}

\makeatletter
\def\blfootnote{\xdef\@thefnmark{}\@footnotetext}
\makeatother

\makeatletter
\def\blfootnote{\xdef\@thefnmark{}\@footnotetext}
\makeatother

\begin{document}

\title{\huge{Companding and Predistortion Techniques for Improved Efficiency and Performance in SWIPT}}

\author{Santiago Fernández, F. Javier L\'opez-Mart\'inez,  Fernando H. Gregorio and Juan Cousseau}

\maketitle
\begin{abstract}
In this work, we analyze how the use of companding techniques, together with digital predistortion (DPD), can be leveraged to improve system efficiency and performance in simultaneous wireless information and power transfer (SWIPT) systems based on power splitting. By taking advantage of the benefits of each of these well-known techniques to mitigate non-linear effects due to power amplifier (PA) and energy harvesting (EH) operation, we illustrate how DPD and companding can be effectively combined to improve the EH efficiency while keeping unalterable the information transfer performance. We establish design criteria that allow the PA to operate in a higher efficiency region so that the reduction in peak-to-average power ratio over the transmitted signal is translated into an increase in the average radiated power and EH efficiency. The performance of DPD and companding techniques is evaluated in a number of scenarios, showing that a combination of both techniques allows to significantly increase the power transfer efficiency in SWIPT systems.
\end{abstract}

\blfootnote{\noindent Manuscript received August 11, 2023; revised February 25 and April 10, 2024. This work was funded in part by grant EMERGIA20-00297 funded by Consejer\'ia de Universidad, Investigaci\'on e Innovaci\'on of Junta de Andaluc\'ia, and in part by grant PID2020-118139RB-I00 funded by MICIU/AEI/10.13039/501100011033. The review of this paper was coordinated by Prof. Junhui Zhao. A preliminary conference version of this work is available in \cite{Companding2021}.}
\blfootnote{\noindent S. Fern\'andez, F. Gregorio and J. Cousseau are with Departamento de Ingenier\'ia El\'ectrica y de Computadoras, UNS-CONICET, Instituto de Investigaciones en Ingenier\'ia El\'ectrica ``Alfredo Desages'' (IIIE), Universidad Nacional del Sur, Bah\'ia Blanca 8000, Argentina. Contact e-mail: $\rm sfernandez@iiie$-$\rm conicet.gob.ar$.}
\blfootnote{\noindent F.J. L{\'o}pez-Mart{\'i}nez is with Dept. Signal Theory, Networking and Communications, Research Centre for Information and Communication Technologies (CITIC-UGR), University of Granada, 18071, Granada, (Spain), and also with the Communications and Signal Processing Lab, Telecommunication Research Institute (TELMA), Universidad de M{\'a}laga, M{\'a}laga, 29010, (Spain).}
\section{Introduction}
The telecommunications industry (TI) is responsible for almost $3$ \% of the total $\rm{CO_2}$ emission worldwide. Some studies indicate that the rate of growth of energy consumption and correspondent $\rm{CO_2}$ footprint is doubling about every 5 years, indicating that TI is probably the fastest growing sector in terms of $\rm{CO_2}$ emissions\cite{Strategies_green_ICT_infrastructure}. 
In this new era of internet of things (IoT) and fifth-generation (5G) communications, it is expected that billions of devices are connected and in continuous operation. {\color{black} In this line, it is imperative to find new efficient ways in which these wireless devices can be powered.}

Plausible mechanisms to allow for a greener and more sustainable operation of IoT devices are based on energy harvesting (EH) techniques so that the energy required for their operation is obtained from sources such as solar power, wind, or radio frequency (RF) waves \cite{libro_EH_Wireless_Comm_Chuang, GreenEnergy}. Among these, the latter enables the simultaneous wireless information and power transfer (SWIPT) paradigm \cite{SWIPT_in_modern_CS} so that RF signals are used {both} for communication and energization. Classically, a dedicated wireless source node called power beacon (PB) generates the radio signals used for EH via wireless power transfer (WPT). In SWIPT, these signals also carry information, and a protocol is required to design the EH and information processing operations in the receiver nodes. The power splitting (PS) protocol \cite{book_IoT_SmartCities,TS_PS_analysis_Perera} is a popular option to implement SWIPT so that the available power at the destination node is split into two parts, one for EH and one for information processing \cite{Survey_SWIPT_CoopRelay_Hossain,Ashraf2021SimultaneousWI}. The PS ratio $\rho$, \textcolor{black}{defined as} the fraction of power dedicated to each of these tasks, can be adjusted to optimize system operation. Since the {\color{black}energy harvester} sensitivity is several tens of dBs worse than information receiver sensitivity, the design of the energy-information splitting protocol is not a trivial task \cite{SWIPT_NovelReceiverDesign_CW}.

One of the key challenges in SWIPT systems based on the PS protocol relies in the use of waveforms that are adequate {both} for information and power transfer. In the literature, there is no clear consensus about the optimal waveform for this task. Some authors claim that signals with a high peak-to-average power ratio (PAPR) \cite{Valenta2015} are desirable to improve {\color{black}the efficiency of the energy harvester node (EHn)}, especially at low powers. In this line, designs in \cite{Clerckx_Fundam_SWIPT_EH_models,Boaventura_Boost_Eff_Unconv_Wave,Clerkx_Wave_Desing_WPT} establish the optimality of high-PAPR signals for RF WPT over a flat-fading channel. With this idea, additional efforts have been made to optimize waveforms with the ultimate goal of improving the efficiency of the rectification process inherent to EH \cite{Collado_ChaoticWave, Santi_RPIC2019}. However, other studies favor the opposite choice \textcolor{black}{of a low PAPR} for waveform design, which is justified as follows: On the one hand, the feasibility of generating high-PAPR waveforms in the transmitter side has been often overlooked. This is relevant, since the overall efficiency of a WPT system is \textit{jointly} determined by the efficiencies of \textit{both} the power signal transmission and rectification methodologies, which correspond to the power amplifier (PA) and the {\color{black}energy harvester}. It turns out that when the PA non-linear effects are properly taken into account \cite{E-t-E_Eff_Fineses}, continuous wave (CW) signals are more efficient than high-PAPR signals because of the excessive energy loss at the PB due to the PA poor energy efficiency. On the other hand, experiments like the one carried out in \cite{E-t-E_Eff_Fineses} show that signals with high PAPR are not well-suited for RF WPT over a flat-fading channel, due to their low average radiated power. Signals with this characteristic make the diodes of the rectifier to turn on, even if the harvester sensitivity is above the signal average input power. However, although {\color{black}EHn} sensitivity is upgraded, the periods of conduction are very short, which ultimately reduces its efficiency. Hence, when all non-linear effects at both the {\color{black}PA} and {\color{black}EHn} are taken into account, high-PAPR signals are not as desirable from {both} an EH perspective and an information transmission viewpoint.

{\color{black} In the literature, different alternatives have been considered for SWIPT, including massive Multiple-Input-Multiple-Output (mMIMO) and mmWave solutions \cite{Zhao_DownlinkHybrid_InfEne_trans_mMIMO,Yang_Throughput_Optim_mMIMO_WPT,Uwaechia_Survey_mW_comm_5G_2020, Hu_Integrated_DataEnergy_comm_Survey_2018}. When it comes to waveform design, and} despite their inherently high PAPR, Orthogonal Frequency Division Multiplexing (OFDM) signals have numerous appealing properties for information transfer, such as immunity to impulsive noise and intersymbol interference, high spectral efficiency and low complexity. \textcolor{black}{Hence, they are the waveform of choice for numerous wireless standards.} Because of their high PAPR, caution must be exercised so that the PA efficiency is not penalized. \textcolor{black}{Recently, the use of Orthogonal-Time-Frequency-Space (OTFS) modulation has been proposed in the context of SWIPT, showing improved performance in the information transfer stages in the presence of high Doppler spreads. However, in low-mobility scenarios (i.e., the usual use case for SWIPT), conventional OFDM performance is expected to match that of OTFS \cite{Liu_OTFS_vs_OFDM} with a lesser complexity.} Since OFDM signals have been used for many years in communication systems, there are several techniques that can be used to improve the efficiency of the PA. For instance, digital predistortion (DPD) techniques \cite{Sezginer2006,He2006,Brihuega2021} are used to pre-compensate the PA non-linear characteristic, in a way that good in-band signal quality and low out-of-band emissions are achieved. These have been recently adapted in SWIPT contexts as a way to reduce the PA non-linearities \cite{Mukherjee2020,Park2021}. Another alternative that can be used to reduce the PAPR of the generated OFDM signals is based on the \textcolor{black}{compansion (compression/expansion) or companding} technique \cite{Red_PAPR_muLaw_Wang_Tjhung,Cui2021,Liu2022}. \textcolor{black}{ Classically, companding trades off PAPR reduction with error performance or throughput degradation \cite{Red_PAPR_muLaw_Wang_Tjhung,Cui2021,Liu2022}. In the context of SWIPT, the use of companding can be anticipated as beneficial from a {\color{black}WPT} perspective, as it effectively reduces the PAPR of the OFDM signal. This mitigates the effects of non-linearities at the transmitter side, and is helpful to improve the efficiency of the \textcolor{black}{EHn}. However, it has to be carefully operated, as it may also affect the process of information transfer. Hence, the trade-off between both power and information transfer in the context of SWIPT when companding is used needs to be explored. While the complexity of companding is considerably lower that the DPD counterpart}, its applicability in the context of waveform design for SWIPT systems has not been explored to the best of our knowledge. 

In this work, the effectiveness of companding and DPD techniques in the operation of SWIPT systems based on {\color{black} PS}\footnote{\textcolor{black}{PS seems the recommended choice for a practical SWIPT receiver over a time-switching (TS) counterpart, as it reportedly provides better performances in terms of throughput and harvested energy when the dissimilar EH and information receiver sensitivities are considered \cite{Clerckx_2022_WIPT_theor_protot_exper}.}} \textcolor{black}{will be analyzed}. Specifically, design criteria will be established so that DPD and companding techniques can be combined into an effective OFDM-based waveform design, with the ultimate goal of improving the end-to-end {\color{black}power transfer} efficiency and achievable rate performance of the system. \textcolor{black}{We focus on a single-antenna scenario to understand the impact of these techniques in SWIPT systems, as a key building block that serves as a first step towards their integration in a plausible MIMO-based solution.} The main contributions of this work can be summarized as follows: (\emph{i}) the use of companding techniques for OFDM waveform design in SWIPT is proposed \textcolor{black}{and analyzed} for the first time; (\emph{ii}) the combination of companding and DPD techniques as useful tools to improve the PA and {\color{black}EHn} efficiencies, even for high-PAPR signals (OFDM), is analyzed; (\emph{iii}) the performance of DPD and companding techniques is evaluated in a number of scenarios of interest, showing noticeable improvements in terms of end-to-end {\color{black}power transfer} efficiency.

The remainder of this paper is structured as follows: the system model for the OFDM-based SWIPT system with PS is described in Section \ref{Sec:SystemModel}, where the end-to-end {\color{black}power transfer} efficiency is defined, together with the key non-linearities associated to the PA. Then, in Section III the information rate-power trade-off under such non-linearities is defined. In Section IV, DPD and companding techniques for efficient SWIPT are introduced, and later integrated into the system design procedure in Section V. Performance evaluation is carried out in Section VI, evaluating the end-to-end {\color{black}power transfer} efficiency, achievable rate and bit error rate over different types of channels. Finally, the main conclusions are summarized in Section VII.
\section{System model} \label{Sec:SystemModel}
\subsection{Power splitting protocol} \label{Sec:PS_protocol}
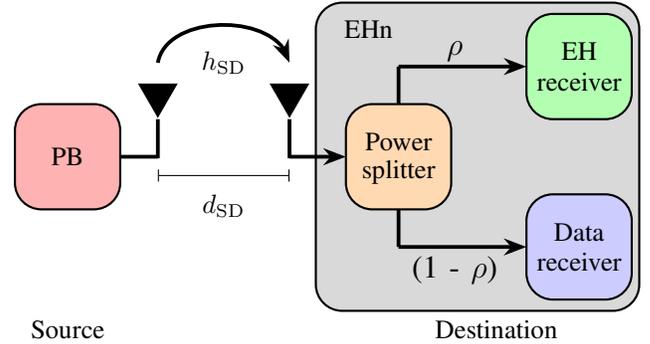
\begin{figure}[t!]
\centering 
\captionsetup{justification=centering}
\begin{tikzpicture}
	\draw[black,fill=red!30,rounded corners=10,thick]
	(0,0) rectangle ++(1.4,1.4);
	\node[centered, centered] at (0.7,0.7) {PB};
	\node[centered, centered] at (0.7,-1.6) {Source};
	
	\draw[black,fill=gray!30,rounded corners=10,thick]
	(4,-1.35) rectangle ++(4.3,4.1);
	\node[centered, centered] at (4.7,2.4) {EHn};
	\node[centered, centered] at (6.4,-1.6) {Destination};
	
	\draw [line width=0.5mm, black ] (1.4,0.7) -- (1.9,0.7) ;
	\draw [line width=0.5mm, black ] (1.9,0.68) -- (1.9,1.2) ;
	\node[isosceles triangle,
	isosceles triangle apex angle=60,
	draw,
	rotate=270,
	fill=black,
	minimum size = 0.35cm] (T2)at (1.9,1.55){};
	
	\draw [line width=0.5mm,-{Stealth[length=3mm]}] (1.9,1.95) to [bend left=80] (3.65,1.95);
	
	\node[centered, centered] at (2.775,1.95) {$h_{\rm SD}$};
	
	\draw [|-|] (1.9,0.45) -- (3.65,0.45);
	\node[centered, centered] at (2.775,0.05) {$d_{\rm SD}$};
	
	\draw [line width=0.5mm, black ] (3.65,0.68) -- (3.65,1.2);
	\draw [line width=0.5mm,-{Stealth[length=3mm]}] (3.65,0.7) -- (4.4,0.7);
	\node[isosceles triangle,
	isosceles triangle apex angle=60,
	draw,
	rotate=270,
	fill=black,
	minimum size = 0.35cm] (T2)at (3.65,1.55){};
	
	\draw[black,fill=orange!30,rounded corners=10,thick]
	(4.4,0) rectangle ++(1.4,1.4);
	\node[centered, centered] at (5.1,0.9) {Power};
	\node[centered, centered] at (5.1,0.5) {splitter};
	
	\draw [line width=0.5mm, black ] (5.1,1.4) -- (5.1,1.9) ;
	\draw [line width=0.5mm,-{Stealth[length=3mm]}] (5.1,1.9) -- (6.8,1.9);
	\node[centered, centered] at (5.85,2.1) {\large $\rho$};
	
	\draw [line width=0.5mm, black ] (5.1,0) -- (5.1,-0.5) ;
	\draw [line width=0.5mm,-{Stealth[length=3mm]}] (5.1,-0.5) -- (6.8,-0.5);
	\node[centered, centered] at (5.85,-0.8) {\large (1 - $\rho$)};
	
	\draw[black,fill=green!30,rounded corners=10,thick]
	(6.8,1.2) rectangle ++(1.4,1.4);
	\node[centered, centered] at (7.5,2.1) {EH};
	\node[centered, centered] at (7.5,1.7) {receiver};
	
	\draw[black,fill=blue!20,rounded corners=10,thick]
	(6.8,-1.2) rectangle ++(1.4,1.4); 	
	\node[centered, centered] at (7.5,-0.3) {Data};
	\node[centered, centered] at (7.5,-0.7) {receiver};	
 \end{tikzpicture}
\caption{\small Power-splitting protocol.}
\label{fig:PS_Protocol}
\vspace{-.4cm} 
\end{figure}

\textcolor{black}{Let us} consider the downlink of a SWIPT system based on PS \cite{MIMO_broad_SWIPT_Zhang_HO}, on which energy and information are wirelessly conveyed through a unique RF signal. \textcolor{black}{At the EHn, the received signal is} split into two streams of adjustable power to separately decode information and harvest energy, \textcolor{black}{as depicted in Fig. \ref{fig:PS_Protocol}. In this figure,} $d_{\rm{SD}}$ is the distance from the PB (source) to the \textcolor{black}{EHn (destination)}, and $h_{\rm{SD}}$ denotes the channel impulse response.

OFDM \cite{Marchetti2009_OFDM,Multicarrier_OFDM_Bahai} \textcolor{black}{is adopted} as the system baseline waveform for simultaneous energy transfer and data transmission, since this type of signals are extensively used in state-of-the-art broadband wireless communications standards, and are also one of the main candidates for beyond-5G and 6G communications systems. OFDM allows to multiplex a set of orthogonal subcarriers, each one modulated with a different modulation index to carry information. Specifically, the incoming bit stream is packed into $b$ bits per symbol to form $X_{\rm k}$, where $b$ is determined by a modulation scheme. Then, $X_{\rm k}$ are processed through an inverse discrete Fourier transform (IDFT), generating the baseband OFDM signal as:
\begin{equation}
    x(n) = \frac{1}{\sqrt{N}} \sum ^{N-1}_{k=0} X_k e^{\frac{j2\pi kn}{N}}.
\end{equation}

After that, the generated digital signal $x(n)$ is processed by a digital-to-analog converter (DAC), obtaining $x(t)$. Finally, the time domain signal is amplified by the PA, obtaining $s(t)$, and transmitted at a certain carrier frequency. \textcolor{black}{To reduce complexity, we assume that the system does not require channel state information (CSI) availability at the transmitter side for its operation.}

One of the key parameters of the OFDM signal is the PAPR, which is defined as: 
\begin{equation}
    {\rm PAPR (dB)}=10\log\left( \frac{\max \{ |x(t)|^2 \}}{\mathbb{E} \{|x(t)|^2} \right),
\end{equation}
\noindent where $\mathbb{E}\{ \cdot \}$ is the expectation operator.
\subsection{\color{black}{End-to-end power transfer efficiency in WPT systems}} \label{Sec:Energy_Eff_WPT}
{\color{black}Power transfer} efficiency is a key aspect in the design of WPT systems, since maximization of the useful power at the input of the {\color{black}EHn} is pursued while keeping the consumed power at the PB within acceptable levels. In order to quantify the WPT system efficiency, the end-to-end {\color{black}energy conversion efficiency or power transfer} efficiency \cite{1G_MobilePowerNetworks_Clerckx} is conventionally used in the literature, as it includes the transmitter efficiency, the effects of the wireless channel, antennas gains, the effects of impedance mismatch of PB and destination node antennas, and the efficiency of the \textcolor{black}{EHn}, as
\begin{equation}
\label{Eq:EtE_Eff}
    \eta_{\rm E\text{-}to\text{-}E} = \frac{P_{\rm DC,ST}}{P_{\rm DC,Tx}} = \underbrace{\frac{P_{\rm RF,Tx}}{P_{\rm DC,Tx}}}_{\eta_1} \underbrace{\frac{P_{\rm RF,Rx}}{P_{\rm RF,Tx}}}_{\eta_2} \underbrace{\frac{P_{\rm DC,Rx}}{P_{\rm RF,Rx}}}_{\eta_3} \underbrace{\frac{P_{\rm DC,ST}}{P_{\rm DC,Rx}}}_{\eta_4},
\end{equation}

\noindent where: $P_{\rm{DC,ST}}$ is the useful power available at the {\color{black}EHn} (to store or use), $P_{\rm{DC,Tx}}$ is the DC power consumed by the PB, $P_{\rm{RF,Tx}}$ is the power radiated by the transmitting antenna, $P_{\rm{RF,Rx}}$ is the power received by the receiver antenna and $P_{\rm{DC,Rx}}$ is the output DC power of the rectifier circuit. Each of these single-stages represents a specific efficiency; for instance, the PB efficiency to convert DC power to RF energy and transmit is represented by $\eta_1$. We note that this stage includes the consumption of \textit{all} devices in the transmitter chain. Usually, the main power consumption in the PB is due to the PA. For this reason, in the sequel we assume that $\eta_1$ is approximately equal to the PA average efficiency.  
The \textcolor{black}{conversion} efficiency $\eta_2$ represents the losses produced by the wireless channel, and chiefly depends on the distance between PB and destination node antennas. The {\color{black}EHn} \textcolor{black}{conversion} efficiency is represented by $\eta_3$ and is related to the capability of the {\color{black}EHn} to convert the incoming RF signals into DC power. 
Finally, if there was a DC-DC converter in the {\color{black}EHn}, its \textcolor{black}{conversion} efficiency would be represented by the term $\eta_4$. Often, the terms $\eta_3$ and $\eta_4$ are grouped into one to represent the ratio of the total amount of power available for storage to the total power received by the {\color{black}EHn}. On the other hand, the terms $\eta_1$ and $\eta_3$ are dependent on the waveform; thus, it is possible to design a certain waveform in order to maximize both efficiencies.
\subsubsection{$\eta_1$ (Power amplifier \textcolor{black}{conversion} efficiency)}
PA efficiency is the most problematic when transmitting high-PAPR signals \cite{E-t-E_Eff_Fineses}, due to the highly non-linear behavior of the PA beyond the saturation point. This implies that when a high-PAPR signal is amplified, its amplitude peaks could saturate the PA, generating distortion and out-of-band radiation due to the output clipping. 

When dealing with high PAPRs, the average input power at the PA needs to be adjusted so that clipping is avoided using an input-backoff (IBO). For instance, setting ${\rm IBO} = {\rm PAPR}$, gives a reasonably-low clipping probability level, and is often used as a rule of thumb. However, the use of an IBO drastically reduces the efficiency of the PA, so a trade-off between efficiency and non-linearity should be established.
 
For the case of considering a class-A PA\footnote{\textcolor{black}{Class-A power amplifiers operate with a constant current, independently of the input signal. Hence, they yield an excellent linearity at the expense of a limited power efficiency \cite{libro_fer_etal}.}} \cite{phdthesis_FerGregorio}, the average power efficiency as a function of PAPR is given by \cite{libro_fer_etal}
\begin{equation}
    \overline{\eta}_{\rm{PA}} = \overline{\eta}_{\rm{PA}_{max}}/{\rm PAPR},
    \label{Eq:PA_Eff}
\end{equation}
\noindent where $\overline{\eta}_{\rm{PA}} = \overline{\eta}_{\rm{PA}_{max}}=50$ \% is only reached for a constant envelope signal $({\rm PAPR}=1)$ with the PA operating at the saturation point \cite{libro_fer_etal}, i.e., ${\rm{IBO}} = 0$ dB. When OFDM signals are used, the PA efficiency is dramatically reduced: for example, for a clipping probability of $10^{-4}$, then ${\rm PAPR}=12$ dB is required with $N=1024$. This means that the PA is operating $12$ dB above the average input power and reaches a power efficiency close to $3$\%. If the PAPR is reduced to $10$ dB, its efficiency is increased to $6$\% \cite{libro_fer_etal}. These poor results in terms of efficiency motivate the development of compensation techniques, such as predistortion and PAPR reduction methods.
\subsubsection{$\eta_3$ (Energy harvester \textcolor{black}{conversion} efficiency)} \label{Sec:Rectenna_Models}
The front-end of an {\color{black}EHn} is equipped with a rectifying antenna, commonly known as rectenna. This device consists of a receiving antenna and a rectifier circuit that converts the energy carried by an RF signal into a DC voltage that can be stored, for example, in a battery. As previously mentioned, {\color{black}EHn} efficiency is defined by the capability of the {\color{black}EHn} to convert the incoming RF signals into DC power. Hence, maximizing RF-to-DC conversion efficiency requires designing efficient rectennas.
  
In WPT, the rectenna can be optimized for the specific operating frequencies, input power level and output load\cite{1G_MobilePowerNetworks_Clerckx}. 
In this line, the input power level defines the topology of the rectifier circuit, since a minimum amount of input power is needed in order to turn on the rectifying devices \textcolor{black}{due to sensitivity}. This implies that it is possible to reach better RF-to-DC conversion efficiencies even with topologies with less number of rectifying devices\cite{OptimumBehavior_WPT_Boaventura_Collado}.  
\subsection{PA non-linear distortion}
Considering a green wireless system, \textcolor{black}{understood as one} using reasonably limited levels of transmitted power, a PA model without memory is assumed in our study. The PA is completely characterized by their AM/AM (amplitude to amplitude) and AM/PM (amplitude to phase) conversions which depend only on the current input signal value. Considering a solid state PA (SSPA), the AM/PM conversion can be neglected and its AM/AM response can be written as
\begin{equation}
    F_a(|x(t)|) = \frac{|x(t)|}{\left[ 1 + \left( \tfrac{|x(t)|}{A_s}\right)^{2p}\right]^{1/2p}},
    \label{eq:PA_AM/AM}
\end{equation}
\noindent where $F_a(\cdot)$ is the AM/AM characteristic, $p$ denotes the smoothness of the transition from linear operation to saturation, and $A_s$ is the saturation output amplitude. For large values of $p$, the SSPA model approaches the  soft limiter (SL) model, described by the simplified expression:
\begin{equation}
s(t) = \left\{ \begin{array}{lcc}
\frac{A_{\rm c}}{\nu} x(t) &   \textrm{for} & |x(t)| \leq \frac{A_{\rm s}}{A_{\rm c}} \nu, \\
A_{\rm s} &   \textrm{for} & |x(t)| > \frac{A_{\rm s}}{A_{\rm c}} \nu,
\end{array}
\right.
\label{Eq:SF_PA}
\end{equation}
\noindent with $A_{\rm s}$ being the PA saturation voltage, $A_{\rm c} = A_{\rm s}/{\mathbb{E}}\left\{ |x(t)|^2\right\}$ the clipping level, and $\nu^2$ the IBO.

Now, if the OFDM signal at the input of the PA is modeled using Gaussian statistics, \textcolor{black}{it is possible to express the output of the memoryless non-linear PA $g[\cdot]$ in a linearized form. As described in \cite{Dardari_NLD}, by virtue of Bussgang theorem, the PA output \textcolor{black}{can be expressed} as}
\begin{equation}
    s(t) = g[x(t)] = K_{\rm L} x(t) + w_{\rm d}(t),
    \label{Eq:Bussgang}
\end{equation}
\noindent \textcolor{black}{where $w_{\rm d}(t)$ is a noise-like non-linear distortion term, modeled as a zero-mean process uncorrelated from $x(t)$, with variance $\sigma_{\rm d}^2$. Similarly, $K_{\rm L}$ is a constant scaling} factor depending on the PA transfer function and its operation point. 
The parameters of the equivalent linear model can be calculated as follows \cite{libro_fer_etal}:
 \begin{gather}
     K_{\rm L} = \frac{\mathbb{E}\left\{ x^*(t)  g[x(t)]\right\}}{\mathbb{E}\left\{ |x(t) |^2 \right\}},
    \label{eq:KL} \\
    \sigma_{\rm d}^2 = \mathbb{E}\left\{ |g[x(t)]|^2 \right\} - |K_{\rm L}|^2 \mathbb{E}\left\{ |x(t) |^2 \right\},
    \label{eq:sigmad}
 \end{gather}
\noindent and the solution of Eqs. (\ref{eq:KL}) and (\ref{eq:sigmad}) for the SL PA model are given by
\begin{gather}
K_{\rm L}(\nu) = \frac{A_{\rm c}}{\nu} \left(1-\exp(-\nu^2)+\frac{\sqrt{\pi}\nu}{2} \mathrm{erfc}(\nu)\right)
 \label{eq:KL_IBO}, \\
\sigma_{\rm d}^2(\nu) = \frac{A_{\rm c}}{\nu} \left( 1-\exp(-\nu^2) - K_{\rm L}^2(\nu) \right),
\label{eq:sigmad_IBO}
\end{gather}
\noindent where $\rm{erfc}(\cdot)$ is the {\it complementary error function}.
\section{Signal model for SWIPT with non-linearities}
\label{Sec:Rate_HarvEner_TradeOff}
\subsection{Formulation}
According to Eq. (\ref{Eq:Bussgang}), the amplification process by the PA introduces an additive distortion term $w_{\rm d}(t)$ with zero mean and variance $\sigma_{\rm d}^2$. The resultant scaling factor $K_{\rm L}$ and the distortion term $w_{\rm d}$ depend on the PA operation point, regulated by the IBO, which is ultimately dependent on the signal PAPR. Defining $\Gamma_{\rm E} = \rho P_{\rm{RF,Tx}}$ and $\Gamma_{\rm I} = (1-\rho) P_{\rm{RF,Tx}}$, the received signals by the energy-constrained destination node after PS, i.e., for EH and information processing, respectively, can be expressed as:
\begin{equation} 
y(t) = \left\{ 
\begin{array}{lc} 
 \sqrt{\Gamma_{\rm E}} h_{\rm SD}(t) *s(t)  + \sqrt{\rho} w_{\rm a}(t) + w_{\rm p}(t), \\  
 \sqrt{\Gamma_{\rm I}} h_{\rm SD}(t) *s(t)  + \sqrt{1-\rho}  w_{\rm a}(t) + w_{\rm p}(t), 
\end{array}
\right. 
\label{Eq:InputSignalPS}
\end{equation}
\noindent with $0 \leq t \leq T$, where $h_{\rm SD}(t)$ is the channel impulse response, $*$ denotes the convolution operation, $w_{\rm a}(t)$ is the complex additive white Gaussian noise (AWGN) introduced by the receive antenna with zero mean and variance $\sigma_{\rm a}^2$, and $w_{\rm p}(t)$ is the complex AWGN introduced by the receiver processing with zero mean and variance $\sigma_{\rm p}^2$. For the sake of notational simplicity, we assume a flat fading channel so that the convolution operation is replaced by a product operation with the single-tap channel coefficient $h_{\rm SD}$. Later, the analysis will be extended to include the effect of multipath propagation.

Taking into account the distortion introduced by the PA in Eq. \eqref{Eq:Bussgang}, the latter expression can be written as:
\begin{equation} 
y(t) = \left\{ 
\begin{array}{lc} 
 \biggr[h_{\rm SD} \sqrt{\Gamma_{\rm E}} \left( K_{\rm L}  x(t) + w_{\rm d}(t) \right)  \\ \quad + \sqrt{\rho} w_{\rm a}(t) + w_{\rm p}(t)\biggr], \\ \\
\biggr[ h_{\rm SD} \sqrt{\Gamma_{\rm I}} \left( K_{\rm L} x(t) + w_{\rm d}(t) \right) \\  \quad + \sqrt{1-\rho}  w_{\rm a}(t) + w_{\rm p}(t)\biggr], \\
\end{array}
\right. 
\end{equation}
\noindent with $0 \leq t \leq T$.
Hence, the received signal destined to information recovery is given by:
\begin{align}
y_I(t) = & \left[ \underbrace{h_{\rm SD} \sqrt{\Gamma_{\rm I}} K_{\rm L} x(t)}_{\rm Desired \ signal} \right. \nonumber \\ &  \left. + \underbrace{h_{\rm SD} \sqrt{\Gamma_{\rm I}} w_{\rm d}(t) + \sqrt{1-\rho}  w_{\rm a}(t) + w_{\rm p}(t)}_{\rm Undesired \ signal} \right],
\end{align}
\noindent where the signal-to-interference plus noise ratio (SINR) can be formulated as follows:
\begin{equation}
{\rm SINR} = \frac{|h_{\rm SD}|^2 \Gamma_{\rm I} K_{\rm L}^2}{|h_{\rm SD}|^2 \Gamma_{\rm I} \sigma_{\rm d}^2 + (1-\rho) \sigma_{\rm a}^2 + \sigma_{\rm p}^2},
\end{equation}
\noindent where we assumed without loss of generality that $\mathbb{E}\{|x|^2\} = 1$, i.e., all transmitted symbols have normalized unit power. Then, an achievable rate per bandwidth unit can be calculated by
\begin{equation}
C = \log_2 \left( 1 + \frac{|h_{\rm SD}|^2 \Gamma_{\rm I} K_{\rm L}^2}{|h_{\rm SD}|^2 \Gamma_{\rm I} \sigma_{\rm d}^2 + (1-\rho) \sigma_{\rm a}^2 + \sigma_{\rm p}^2} \right).
\label{Eq:C_PS}
\end{equation}
Similarly, the received signal intended for EH is given by
\begin{align}
y_H(t) = & \left[ h_{\rm SD} \sqrt{\Gamma_{\rm E}}  K_{\rm L}  x(t)  \right. \nonumber \\ &\left.+ h_{\rm SD} \sqrt{\Gamma_{\rm E}} w_{\rm d}(t) + \sqrt{\rho} w_{\rm a}(t) + w_{\rm p}(t)\right],
\end{align}
\noindent where now all terms contribute to energy harvesting. The total harvested energy is expressed by
{\color{black}
\begin{equation}
H_{\rm E} = \eta_{3}\left(|h_{\rm SD}|^2 \Gamma_{\rm E} (K_{\rm L}^2 + \sigma_{\rm d}^2) + \rho  \sigma_{\rm a}^2 + \sigma_{\rm p}^2 \right) T,
\label{Eq:E_PS}
\end{equation}
}
\noindent where \textcolor{black}{$\eta_{3}(\cdot)$ is the function describing the conversion efficiency of the EHn, which is in general non-linear\footnote{\textcolor{black}{In the literature, the case of linear behavior for EHn is often used for the sake of simplicity and benchmarking purposes. In such a case, $\eta_3(P_{\rm in})=\eta_{3,{\rm L}}P_{\rm in}$, where $\eta_{3,{\rm L}}$ is a constant value.}}.} Since power is defined as energy per unit time, the power available for EH can be expressed by
{\color{black}
\begin{equation}
    H_{\rm P} = H_{\rm E}/T,
    \label{Eq:P_EH}
\end{equation}
}
\textcolor{black}{
        Based on Eqs. \eqref{Eq:E_PS} and \eqref{Eq:P_EH} and assuming a normalized symbol period $T$, we define the metric $H_{\rm P/E}$ making explicit that the harvested energy metric and the harvested power metric may be used interchangeably \cite{Chen_EH_Principles_Theory}. The unit of this metric is joule when measuring harvested energy, or watt when measuring harvested power but, for simplicity, $H_{\rm P/E}$ \textcolor{black}{is normalized} to the maximum transmit power $P$, so that it is a dimensionless metric.
        }
\subsection{Simulations and observations} 
\label{SubSec:SignalModelSWIPT_nonLinear_SimulObserv}
\begin{figure}[t!]
\centering
  \subfloat[Rate $C$ as a function of $\rho$.] {\includegraphics[width=1\columnwidth]{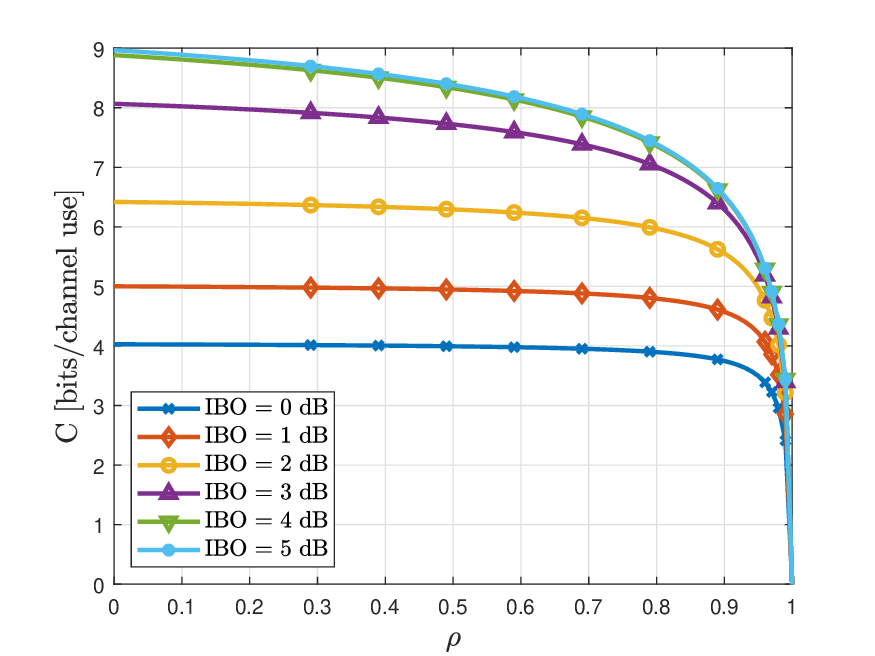}
  \label{fig:Rate_rho_PS}}
  \\ 
  \subfloat[Power/energy harvested $H_{\rm P/E}$ as a function of $\rho$.] 
  {\includegraphics[width=1\columnwidth]{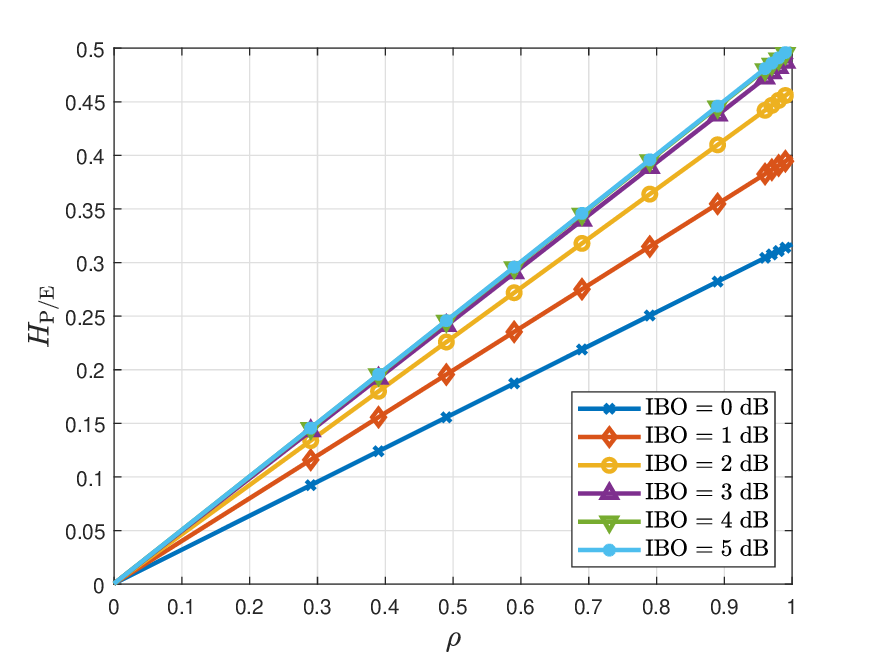}
  \label{fig:Power_rho_PS}} 
  \caption{\small Rate and power/energy harvested functions as a function of $\rho$. With $P_{\rm{RF,Tx}} = 1$, $h_{\rm SD} = 1$, $\eta_{3,{\rm L}} = 0.5$ and $\sigma_{\rm a}^2 = \sigma_{\rm p}^2 = 10^{-3}$.} \label{Fig:Rate_Power_rho_PS}
\end{figure}

In order to better reflect the inherent trade-offs between PA non-linearity, IBO, PAPR and PS factor $\rho$ for the information rate and {\color{black}power transfer} efficiency metrics, some illustrative simulations \textcolor{black}{are provided}. \textcolor{black}{Specifically, Eqs. (\ref{Eq:C_PS}) and metric $H_{\rm P/E}$ are used with typical parameters (see figure caption) to generate Fig. \ref{Fig:Rate_Power_rho_PS};} this is useful to infer that high levels of energy can be harvested with high values of $\rho$, at the expense of a reduction in the information rate. We also see that \textit{both} information rate and harvested power can be improved by a proper design of IBO.
\section{Signal Processing Techniques for Efficient WPT} 
\label{Sec:Tech_Improv_EE}
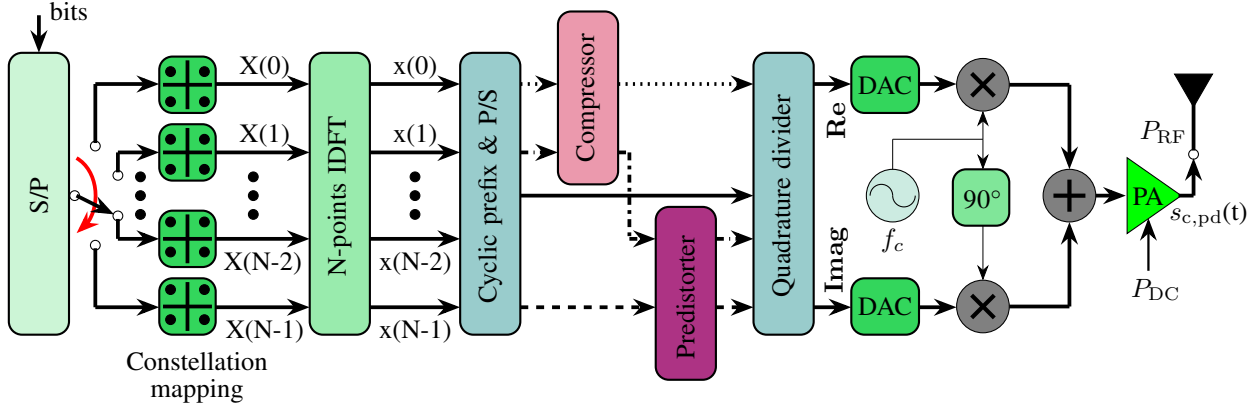
\begin{figure*}[ht!]
\centering 
\begin{tikzpicture}
 	\let\radius\undefined
 	\newlength{\radius}
 	\setlength{\radius}{0.65mm}
 	
 	\draw [line width=0.5mm,-{Stealth[length=3mm]}] (0.4,2.4) -- (0.4,1.9);
 	\node[centered, align=center] at (0.8,2.45) {bits};
 	
 	\draw[black,fill=green!80!blue!20,rounded corners=5,thick]
 	(0,-1.85) rectangle ++(0.8,3.75);
 	\node[centered, align=center,rotate=90] at (0.4,0) {S/P};
 	\coordinate (p1) at (0.8+0.075,0);
 	\draw (p1) circle (\radius);
 	
 	\draw [red,line width=0.5mm,-{Stealth[length=3mm]}] (0.9,0.5) to [bend left=60] (0.9,-0.5); 	
 	
 	\draw [line width=0.5mm,-{Stealth[length=3mm]}] (0.9,0) -- (1.4,-0.25);
 	
 	\coordinate (p2) at (1.45,-0.27);
 	\draw (p2) circle (\radius) ;
 	
 	\coordinate (p3) at (1.45,0.27);
 	\draw (p3) circle (\radius) ;
 	
 	\coordinate (p4) at (1.15,0.66);
 	\draw (p4) circle (\radius) ;
 	
 	\coordinate (p5) at (1.15,-0.66);
 	\draw (p5) circle (\radius) ;
 	
 	\draw[black,fill=green!80!blue!80,rounded corners=5,thick]
 	(2,1.1) rectangle ++(0.75,0.75);
 	\draw [line width=0.5mm, black ] (2.05,1.475) -- (2.7,1.475);
 	\draw [line width=0.5mm, black ] (2.375,1.15) -- (2.375,1.8);
 	\draw [black, fill=black] (2.15,1.65) circle (\radius);
 	\draw [black, fill=black] (2.6,1.65) circle (\radius);
 	\draw [black, fill=black] (2.15,1.25) circle (\radius);
 	\draw [black, fill=black] (2.6,1.25) circle (\radius);
 	
 	\draw[black,fill=green!80!blue!80,rounded corners=5,thick]
 	(2,0.2) rectangle ++(0.75,0.75);
 	\draw [line width=0.5mm, black ] (2.05,0.575) -- (2.7,0.575);
 	\draw [line width=0.5mm, black ] (2.375,0.25) -- (2.375,0.9);
 	\draw [black, fill=black] (2.15,0.75) circle (\radius);
 	\draw [black, fill=black] (2.6,0.75) circle (\radius);
 	\draw [black, fill=black] (2.15,0.35) circle (\radius);
 	\draw [black, fill=black] (2.6,0.35) circle (\radius);
 	
 	\draw[black,fill=green!80!blue!80,rounded corners=5,thick]
 	(2,-0.95) rectangle ++(0.75,0.75);
 	\draw [line width=0.5mm, black ] (2.05,-0.575) -- (2.7,-0.575);
 	\draw [line width=0.5mm, black ] (2.375,-0.25) -- (2.375,-0.9);
 	\draw [black, fill=black] (2.15,-0.75) circle (\radius) ;
 	\draw [black, fill=black] (2.6,-0.75) circle (\radius);
 	\draw [black, fill=black] (2.15,-0.35) circle (\radius);
 	\draw [black, fill=black] (2.6,-0.35) circle (\radius);
 	
 	\draw[black,fill=green!80!blue!80,rounded corners=5,thick]
 	(2,-1.85) rectangle ++(0.75,0.75);
 	\draw [line width=0.5mm, black ] (2.05,-1.475) -- (2.7,-1.475);
 	\draw [line width=0.5mm, black ] (2.375,-1.15) -- (2.375,-1.8);
 	\draw [black, fill=black] (2.15,-1.65) circle (\radius) ;
 	\draw [black, fill=black] (2.6,-1.65) circle (\radius);
 	\draw [black, fill=black] (2.15,-1.25) circle (\radius) ;
 	\draw [black, fill=black] (2.6,-1.25) circle (\radius);
 	
 	\draw [black, fill=black] (1.75,0.25) circle (\radius);
 	\draw [black, fill=black] (1.75,0) circle (\radius);
 	\draw [black, fill=black] (1.75,-0.25) circle (\radius);
 	
 	\draw [line width=0.5mm,-{Stealth[length=3mm]}] (1.15,1.485) -- (2,1.485);
 	\draw [line width=0.5mm, black ] (1.15,0.7) -- (1.15,1.485);

	\draw [line width=0.5mm,-{Stealth[length=3mm]}] (1.15,-1.485) -- (2,-1.485);
	\draw [line width=0.5mm, black ] (1.15,-0.7) -- (1.15,-1.485);
	
	\draw [line width=0.5mm, black ] (1.45, 0.31) --(1.45,0.575); 	
	\draw [line width=0.5mm,-{Stealth[length=3mm]}] (1.45,0.575) -- (2,0.575);
	 	
 	\draw [line width=0.5mm, black ] (1.45, -0.31) --(1.45,-0.575); 	
 	\draw [line width=0.5mm,-{Stealth[length=3mm]}] (1.45,-0.575) -- (2,-0.575);
 	
 	\draw [line width=0.5mm,-{Stealth[length=3mm]}] (2.75,1.485) -- (4,1.485);
 	
 	\draw [line width=0.5mm,-{Stealth[length=3mm]}] (2.75,0.575) -- (4,0.575);
 	
	\draw [line width=0.5mm,-{Stealth[length=3mm]}] (2.75,-0.575) -- (4,-0.575);

 	\draw [line width=0.5mm,-{Stealth[length=3mm]}] (2.75,-1.485) -- (4,-1.485);
 	
 	\node[centered, align=center] at (3.4,1.7) {X(0)};
 	\node[centered, align=center] at (3.4,0.8) {X(1)};
 	\node[centered, align=center] at (3.375,-0.9) {X(N-2)};
 	\node[centered, align=center] at (3.4,-1.8) {X(N-1)};
 	
 	\draw [black, fill=black] (3.25,0.25) circle (\radius);
 	\draw [black, fill=black] (3.25,0) circle (\radius);
 	\draw [black, fill=black] (3.25,-0.25) circle (\radius);
 	
 	\node[centered, align=center] at (2.5,-2.2) {Constellation};
 	\node[centered, align=center] at (2.5,-2.6) {mapping};

 	\draw[black,fill=green!80!blue!40,rounded corners=5,thick]
 	(4,-1.85) rectangle ++(0.8,3.75);
 	\node[centered, align=center,rotate=90] at (4.4,0) {N-points IDFT};
 	
 	\draw [line width=0.5mm,-{Stealth[length=3mm]}] (4.8,1.485) -- (6,1.485);
 	\draw [line width=0.5mm,-{Stealth[length=3mm]}] (4.8,0.575) -- (6,0.575);
 	\draw [line width=0.5mm,-{Stealth[length=3mm]}] (4.8,-0.575) -- (6,-0.575);
 	\draw [line width=0.5mm,-{Stealth[length=3mm]}] (4.8,-1.485) -- (6,-1.485);
 	
 	\node[centered, align=center] at (5.4,1.7) {x(0)};
 	\node[centered, align=center] at (5.4,0.8) {x(1)};
 	\node[centered, align=center] at (5.4,-0.9) {x(N-2)};
 	\node[centered, align=center] at (5.4,-1.8) {x(N-1)};
 	
 	\draw [black, fill=black] (5.4,0.25) circle (\radius);
 	\draw [black, fill=black] (5.4,0) circle (\radius);
 	\draw [black, fill=black] (5.4,-0.25) circle (\radius);
 	
 	\draw[black,fill=green!50!blue!40,rounded corners=5,thick]
 	(6,-1.85) rectangle ++(0.8,3.75);
 	\node[centered, align=center,rotate=90] at (6.4,0) {Cyclic prefix \& P/S};
 	
 	\draw[-{Stealth[length=3mm]},thick,black,dotted,line width=0.5mm] (6.8,1.485) -- (7.3,1.485);
 	
 	\draw[-{Stealth[length=3mm]},thin,black,dash dot,line width=0.5mm] (6.8,0.575) -- (7.3,0.575);
 	
 	\draw[-{Stealth[length=3mm]},thick,black,line width=0.5mm] (6.8,0) -- (9.9,0);
 	
 	\draw[-{Stealth[length=3mm]},thick,black,dashed,line width=0.5mm] (6.8,-1.485) -- (8.6,-1.485);
 	
 	\draw[black,fill=red!80!blue!40,rounded corners=5,thick]
 	(7.3,0.15) rectangle ++(0.8,2.3);
 	\node[centered, align=center,rotate=90] at (7.7,1.3) {Compressor};
 	
 	\draw [line width=0.5mm, black, dash dot ] (8.1,0.575) -- (8.25,0.575);	
 	\draw [line width=0.5mm, black, dash dot ] (8.25,0.575) -- (8.25,-0.575);
 	\draw[-{Stealth[length=3mm]},thin,black,dash dot,line width=0.5mm] (8.25,-0.575) -- (8.6,-0.575);
 	
 	\draw[-{Stealth[length=3mm]},thick,black,dotted,line width=0.5mm] (8.1,1.485) -- (9.9,1.485);
 	
 	\draw[black,fill=red!60!blue!80,rounded corners=5,thick]
 	(8.6,-0.15) rectangle ++(0.8,-2.3);
 	\node[centered, align=center,rotate=90] at (9,-1.35) {Predistorter};
 	
 	\draw[-{Stealth[length=3mm]},thin,black,dash dot,line width=0.5mm] (9.4,-0.575) -- (9.9,-0.575);
 	
 	\draw[-{Stealth[length=3mm]},thick,black,dashed,line width=0.5mm] (9.4,-1.485) -- (9.9,-1.485);
 	
 	\draw[black,fill=green!50!blue!40,rounded corners=5,thick]
 	(9.9,-1.85) rectangle ++(0.8,3.75);
	\node[centered, align=center, rotate=90] at (10.3,0) {Quadrature divider};	
	
	\draw[-{Stealth[length=3mm]},thick,black,line width=0.5mm] (10.7,1.485) -- (11.2,1.485);
	\draw[-{Stealth[length=3mm]},thick,black,line width=0.5mm] (10.7,-1.485) -- (11.2,-1.485);
	
	\node[centered, align=center, rotate=90] at (11,1) {$\mathbf{Re}$};
	\node[centered, align=center, rotate=90] at (11,-0.85) {$\mathbf{Imag}$};	
		
	 \draw[black,fill=green!80!blue!80,rounded corners=5,thick]
	 (11.2,1.1) rectangle ++(0.9,0.75);
	 \node[centered, align=center] at (11.65,1.475) {DAC};
	 
	 \draw[black,fill=green!80!blue!80,rounded corners=5,thick]
	 (11.2,-1.85) rectangle ++(0.9,0.75);
	 \node[centered, align=center] at (11.65,-1.475) {DAC};
	
	\draw[-{Stealth[length=3mm]},thick,black,line width=0.5mm] (12.1,1.485) -- (12.6,1.485);
	\draw[-{Stealth[length=3mm]},thick,black,line width=0.5mm] (12.1,-1.485) -- (12.6,-1.485);
	 
	\draw [black,fill=green!60!blue!20] (11.75,0) circle (3.5mm);
	
	\draw (11.75,0) arc(0:180:1.5mm) ;
	\draw (11.75,0) arc(-180:0:1.5mm) ;
	\node[centered, align=center] at (11.75,-0.6) {$f_c$};
	\draw [line width=0.1mm, black ] (11.75,0.35) -- (11.75,0.75);
	\draw [line width=0.1mm, black ] (11.75,0.75) -- (12.95,0.75);
	\draw[-{Stealth[length=2mm]},thick,black,line width=0.1mm] (12.95,0.75) -- (12.95,0.35);
	\draw[-{Stealth[length=2mm]},thick,black,line width=0.1mm] (12.95,0.75) -- (12.95,1.125);
	\draw[-{Stealth[length=2mm]},thick,black,line width=0.1mm] (12.95,-0.35) -- (12.95,-1.125);
	 
	\draw [black,fill=gray!100] (12.95,1.475) circle (3.5mm); 	
	\node[centered, align=center,rotate = 45] at (12.95,1.475) {\Huge +};
		
 	\draw [black,fill=gray!100] (12.95,-1.475) circle (3.5mm);
 	\node[centered, align=center,rotate = 45] at (12.95,-1.475) {\Huge +};
 	
 	\draw[black,fill=green!80!blue!50,rounded corners=5,thick]
 	(12.55,-0.4) rectangle ++(0.75,0.75);
 	\node[centered, align=center] at (12.95,-0.05) {90$^\circ$};
 	
 	\draw [line width=0.5mm, black ] (13.3,1.475) -- (14.1,1.475);
 	\draw[-{Stealth[length=3mm]},thick,black,line width=0.5mm] (14.1,1.475) -- (14.1,0.35);
 	
 	\draw [line width=0.5mm, black ] (13.3,-1.475) -- (14.1,-1.475);
 	\draw[-{Stealth[length=3mm]},thick,black,line width=0.5mm] (14.1,-1.475) -- (14.1,-0.35);
 	
 	\draw [black,fill=gray!100] (14.1,0) circle (3.5mm);
 	\node[centered, align=center] at (14.1,0) {\Huge +};
 	
 	\draw[-{Stealth[length=3mm]},thick,black,line width=0.5mm] (14.45,0) -- (14.85,0);
 	
 	\node[isosceles triangle,
 	isosceles triangle apex angle=70,
 	draw,
 	rotate=0,
 	fill=green,
 	minimum size = 0.72cm] (T2)at (15.1,0){};
 	\node[centered, align=center] at (15.15,0) {PA};
 	
 	\node[centered, align=center] at (15.25,-1.25) {$P_{\rm DC}$};
 	\draw[-{Stealth[length=3mm]},thick,black,line width=0.3mm] (15.15,-1) -- (15.15,-0.3);
 	\node[centered, align=center] at (15.95,-0.25) {$s_{\rm c,pd}$(t)};

 	\draw [line width=0.5mm, black ] (15.55,0) -- (15.75,0);
 	\draw[-{Stealth[length=3mm]},thick,black,line width=0.5mm] (15.75,0) -- (15.75,0.5);
 	
 	\coordinate (pAnt) at (15.75,0.55);
 	\draw (pAnt) circle (\radius) ;
 	
 	\draw [line width=0.5mm, black ] (15.75,0.6) -- (15.75,1.5);
 	
 	\node[isosceles triangle,
 	isosceles triangle apex angle=60,
 	draw,
 	rotate=270,
 	fill=black,
 	minimum size = 0.35cm] (T2)at (15.75,1.5){};
 	
 	\node[centered, align=center] at (15.35,0.85) {$P_{\rm RF}$};
 	
 	
 	
 	
\end{tikzpicture}
\caption{\small Block diagram of the proposed OFDM transmitter. \textcolor{black}{Solid, dashed, dotted, and dashdotted lines correspond to the cases of baseline OFDM, DPD-only, companding only, and DPD+companding, respectively. }}
\label{fig:OFDM_Companding_PreDistorter_Block_Diagram_Alternative}
\end{figure*}

The efficiency of the PA is governed by its operating point, which needs to be cleverly adjusted to allow for amplification of the input signal with a reasonable distortion. Several techniques can be implemented to improve efficiency, such as PA dynamic range extension and PAPR reduction. Specifically, dynamic range extension can be obtained by using a {\color{black}DPD} technique before the PA. For PAPR reduction, the use of a companding technique is evaluated. A block diagram of the proposed OFDM transmitter is depicted in Fig. \ref{fig:OFDM_Companding_PreDistorter_Block_Diagram_Alternative}, where the DPD and companding blocks are included. Note that the block diagram allows for the use of DPD and companding techniques alone, and combined. 
\subsection{Digital predistortion technique}
\begin{figure}[t!]
\centering
  \subfloat[Transfer functions of the PA, the predistorter and the linearized result.] {\includegraphics[width=1\columnwidth]{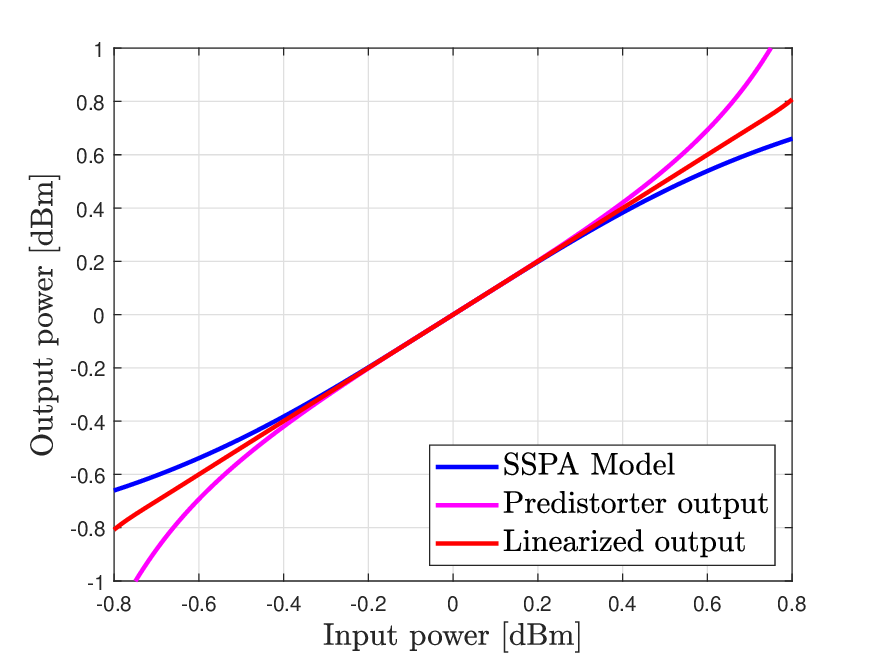}\label{fig:LinearizedPA}} \\
  \subfloat[Block diagram of the predistortion process. DPD: digital predistorter with $f_{\rm PA}^{-1}(\cdot)$ characteristic, DAC: digital-to-analog converter. PA: power amplifier with $f_{\rm PA}(\cdot)$.] {
  \begin{tikzpicture}
	\draw [line width=0.5mm,-{Stealth[length=3mm]}] (0.7,2.4) -- (0.7,1.4);
	\node[centered, align=center] at (1.15,2) {u(n)};
	
	\draw[black,fill=red!30,rounded corners=10,thick]
	(0,0) rectangle ++(1.4,1.4);
	\node[centered, centered] at (0.7,0.7) {$f_{\rm PA}^{-1}(\cdot)$};
	
	\draw [line width=0.5mm, black ] (0.7,0) -- (0.7,-0.7);
	\draw [line width=0.5mm,-{Stealth[length=3mm]}] (0.7,-0.7) -- (1.4,-0.7);
	\draw[black,fill=green!30,rounded corners=10,thick]
	(1.4,-1.4) rectangle ++(1.4,1.4);
	\draw [line width=0.5mm, black ] (2.1,-0.1) -- (2.1,-1.3);
	\draw [line width=0.5mm, black ] (1.5,-0.7) -- (2.7,-0.7);
	
	\draw [line width=0.3mm, black ] (2.1,-0.7) -- (2.3,-0.7)
	;
	\foreach \j in {1,...,6}
	\draw [line width=0.3mm, black ] (1.5+\j/5-0.2,-1.3+\j/5-0.2) -- (1.7+\j/5-0.2,-1.3+\j/5-0.2);
		
	\foreach \j in {1,...,6}
	\draw [line width=0.3mm, black ] (1.7+\j/5-0.2,-1.3+\j/5-0.2) -- (1.7+\j/5-0.2,-1.1+\j/5-0.2)	
	;
	
	\draw [line width=0.5mm,-{Stealth[length=3mm]}] (2.8,-0.7) -- (3.8,-0.7);
	\draw[black,fill=green!70!blue!50,rounded corners=10,thick]
	(3.8,-1.4) rectangle ++(1.4,1.4);
	
	\draw [black,fill=green!100!blue!50] (4.5,-0.7) circle (4.5mm);
	
	\draw (4.5,-0.7) arc(0:180:1.5mm) ;
	\draw (4.5,-0.7) arc(-180:0:1.5mm) ;
	
	\draw [line width=0.5mm, black ] (4.5,0) -- (4.5,0.7);
	\draw [line width=0.5mm,-{Stealth[length=3mm]}] (4.5,0.7) -- (5.7,0.7);
	\node[isosceles triangle,
	isosceles triangle apex angle=65,
	draw,
	rotate=0,
	fill=red!30,
	minimum size = 1.5cm] (T2)at (6.2,0.7){};
	\node[centered, align=center] at (6.3,0.7) {$f_{\rm PA}(\cdot)$};
	
	\draw [line width=0.5mm,-{Stealth[length=3mm]}] (7.2,0.7) -- (8.2,0.7);
	\node[centered, align=center] at (7.6,1) {s(t)};
	
	\node[centered, align=center] at (0.15,-0.25) {DPD};
	\node[centered, align=center] at (2.1,-1.65) {DAC};	
	\node[centered, align=center] at (4.5,-1.65) {Upconverter};		
	\node[centered, align=center] at (6.65,0) {PA};		
\end{tikzpicture} 
\label{fig:BlockDiagram_DPD_PA}} \caption{\small Principles of baseband predistortion technique. }
\label{Fig:Principles_Predistortion}
\end{figure}
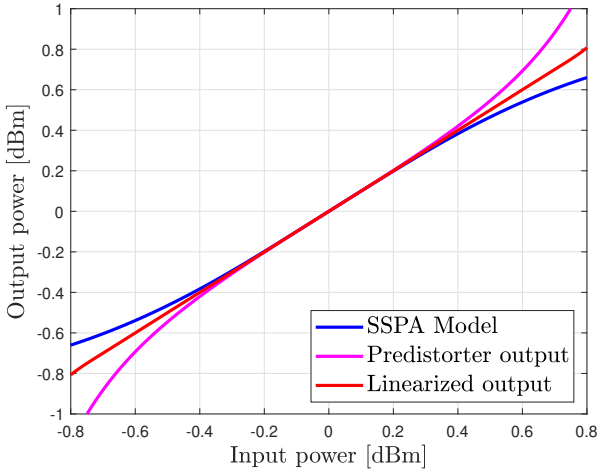
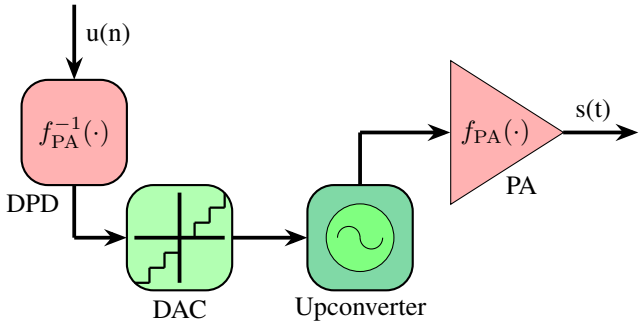

{\color{black} The implementation of DPD is intended to improve the PA linearity, and increase its operation range and energy efficiency. 
This technique allows the reduction of the IBO, which reduces the PA power consumption. DPD} uses a non-linear device located before the non-linear PA, with a transfer function modeling the inverse behavior of the PA. Thanks to this device, it is possible to obtain an almost linearly amplified signal at the output of the system\cite{libro_fer_etal}. The linearization basic principles are depicted in Fig. \ref{fig:LinearizedPA}, and a block diagram describing the principle of the  baseband predistortion process is ilustrated in Fig. \ref{fig:BlockDiagram_DPD_PA}.

{\color{black}A correct estimation of the amplifier response is required to achieve good predistorter performance.
However, the implementation of the inverse of a non-linear model such as the SSPA introduced in \eqref{eq:PA_AM/AM} is a complex task. A practical approach relies on using a polynomial fit to capture the non-linear characteristics of the PA. 
In a first instance, a training sequence with sufficient dynamic range is used to expose the non-linearity feature. Then, a sampled version of the RF output of the PA \cite{Predistortion_Yuanbin} is used in order to obtain these non-linear parameters and to construct a polynomial that describes the behavior of the PA with the desired accuracy. Thus, assuming a memoryless PA and considering only the transfer function of the AM/AM characteristic in \eqref{eq:PA_AM/AM}, the static non-linearity can be modeled using a polynomial as follows
\begin{equation}
	s(t) = {{F}_{a}}[x(t)] \approx \sum^P_{{i=0}}a_{i}|x(t)|^i,
	\label{coefPA}
\end{equation}
\noindent where $x(t)$ is the input signal, $s(t)$ is the output signal of the PA and $\{a_{i} \}^P_{i=0}$ are the polynomial coefficients, where $P$ is the order of the polynomial. 

Through the sampled version of the PA output and using well-known linear parameter estimation methods such as least squares (LS), least mean squares (LMS), or gradient descent algorithm (GD) \cite{PD_tutorial, PD2}, the inverse function of the PA can be modeled by means of a new polynomial.
The output signal of the inverse model $F_{a}^{-1}(\cdot)$ of the PA can be expressed by
\begin{equation}
	x_t(t) = F_{a}^{-1}(s_t(t)) \approx \sum^Q_{j=0}\beta_j s_t^j(t),    
\end{equation}
\noindent where $s_t(t)$ is the output of the PA obtained through the training sequence $x_t(t)$ and $Q$ is the order of the polynomial.

The polynomial orders are selected to obtain a reasonable linearization capability, and also taking into account the numerical stability of the identification algorithm, as high-order polynomials are prone to  oscillatory behaviors and lead to low quality data fits and numerical errors.

Using the sampled versions of $x_t(t)$ y $s_t(t)$, i.e., $x_t[n]$ y $s_t[n]$, the following ancillary signal vectors are defined: $\boldsymbol{\rm{x}}_t = [x_t[1], x_t[2], \cdots x_t[N]]^T$ and  $\boldsymbol{\rm{s}}_t = [s_t[1], s_t[2], \cdots , s_t[N]]^T$, of dimensions $N\times 1$, where $()^T$ denotes the transpose operator.
From these vectors, the {Vandermonde} matrices of dimensions $N \times P$ are built, which are respectively determined by $\Bar{\boldsymbol{\rm{X}}}_t = [\textbf{1}, \boldsymbol{\rm{X}}_t, \boldsymbol{\rm{X}}_t^2, \cdots , \boldsymbol{\rm{X}}_t^P]$ and $\Bar{\boldsymbol{\rm{S}}}_t = [\textbf{1}, \boldsymbol{\rm{S}}_t, \boldsymbol{\rm{S}}_t^2, \cdots , \boldsymbol{\rm{S}}_t^P]$.
 
Using for instance the LS approach, the coefficients $\beta_j$ can be obtained as
\begin{equation}
	\hat{ \boldsymbol{\beta}} = [\Bar{\boldsymbol{\rm{S}}}^{\dagger}_t\Bar{\boldsymbol{\rm{S}}}_t]^{-1} \Bar{\boldsymbol{\rm{S}}}^{\dagger}_t \Bar{\boldsymbol{\rm{X}}}_t,
\end{equation}
\noindent where $\hat{ \boldsymbol{\beta}} = [\beta_0, \beta_1, ..., \beta_Q]$ are the estimated coefficients of $F_{a}^{-1}(\cdot)$ \cite{Predistortion_Yuanbin}.

Finally, the polynomial predistortion function $P_{\rm d}(x)$ of order $Q$  used to linearize the static non-linearity \cite{FER_Split_Predistortion} is given by
\begin{align}
	\hat{P}_{\rm d}[x(t)] \approx & F_{a}^{-1}(s_t(t))  \nonumber \\ = &\sum^Q_{j=0}\beta_j a_j x(t)^j = \sum^Q_{j=0}b_j |x(t)|^j,
	\label{coefPD}
\end{align}
\noindent where $b_j = \beta_j a_j$.

Finally, the output of the predistorter-amplifier cascade can be written as
\begin{equation}
	s(t) \approx \sum^P_{i=0}a_i \left( \sum^Q_{j=0}b_j|x(t)|^j \right)^i.
\end{equation}}
\subsection{Companding technique}
The use of companding techniques for signal PAPR reduction \textcolor{black}{is now analyzed}. Even though different companding techniques \textcolor{black}{were preliminarily evaluated}, $\mu$-law companding \cite{Red_PAPR_muLaw_Wang_Tjhung} \textcolor{black}{was adopted} since it provides satisfactory results with a low-complexity implementation.

The $\mu$-law companding is a popular and robust quantization technique originated in the context of speech signals, used to reduce the dynamic range and avoid quantization distortion. Speech signals are largely made up of small amplitude samples, as these are the most important for speech perception and also the most likely. In contrast, larger amplitude values are not as common and have a low probability of occurrence. This probability of occurrence of high and low amplitudes can be extrapolated to an OFDM signal: for a large number of subcarriers, the OFDM signal in the time domain is well-described by a complex Gaussian distribution. In other words, the highest values have a low probability of occurrence while the probability of occurrence of low and medium values is much larger.

The compressing and expanding functions associated to the $\mu$-law companding technique are described by the following expressions 
\begin{align}
    F_{\rm C}(x) = & \frac{A  \ln \left(1 + \frac{\mu |x|}{A}  \right)}{\ln (1 + \mu)} \textrm{sgn}(x),
    \label{Eq:companding} \\
    F_{\rm C}^{-1}(x) = & \frac{\left( \exp{\left(\frac{|x|\ln(1+\mu)}{A}\right) -1} \right)A}{\mu} \textrm{sgn}(x),
    \label{Eq:decompanding}
\end{align}
\noindent where $A$ specifies the peak magnitude of the input data sequence, $\mu$ is the compression parameter \cite{libro_DigitalCommunications_Sklar}, $\ln$ is the natural logarithm, and $\textrm{sgn}(\cdot)$ is the sign function. 
\section{WPT predistortion and companding design}
\label{Sec:ParametersDesign}
In this section, the design strategies for DPD and companding stages \textcolor{black}{are discussed}. While the design of the former follows a similar rationale as classical DPD architectures non-specific of WPT, the design of the companding stage requires a more detailed formulation and justification.
\subsection{Predistortion technique}
\label{Sec:Optimal_IBO_factor}
For the case of the DPD technique, thanks to the extension of the PA operation range it is possible to reduce the IBO levels and reach better levels of PA efficiency. Hence, the goal is to design the IBO so that the non-linear effects of the PA are minimized, and the PA operates in a region of higher efficiency. 

\textcolor{black}{For a given PA, i.e., smoothness factor $p$ and saturation output amplitude $A_s$, the steps to obtain a target IBO reduction factor are described as follows:} \textcolor{black}{First, the approximation order $P$ for the polynomial approximation in (20) is set. Once this is done, the coefficients $\{a_{i} \}^P_{i=0}$ are computed using LS. Second, the approximation order $Q$ for the polynomial approximation in (23) is set, and the coefficients $\{b_{j} \}^Q_{j=0}$ are computed accordingly by LS algorithm. Finally, the IBO reduction factor is designed with the criterion that the error vector magnitude (EVM) of the output signal when using predistorter is equal to the EVM of the output signal when no predistorter is used.
This series of steps are summarized in Algorithm \ref{alg:LSP}.}
\begin{algorithm}[t!]
	\caption{\footnotesize Digital predistorter stage design}\label{alg:LSP}
	\begin{algorithmic}[1]
		\footnotesize
		\STATE Define approximation order $P$ for the polynomial approximation in \eqref{coefPA}. 
        \STATE Calculate the coefficients $\{a_{i} \}^P_{i=0}$ by LS.%
        \STATE Define approximation order $Q$ for the polynomial approximation in \eqref{coefPD}.%
        \STATE Calculate the coefficients $\{b_{j} \}^Q_{j=0}$ by LS.%
		\STATE Choose IBO reduction factor such that ${\rm EVM_{DPD}}={\rm EVM_{no DPD}}$.
		\end{algorithmic}
\end{algorithm}

\begin{figure}[t!]
\centering
  \subfloat[PA efficiency.] {\includegraphics[width=1\columnwidth]{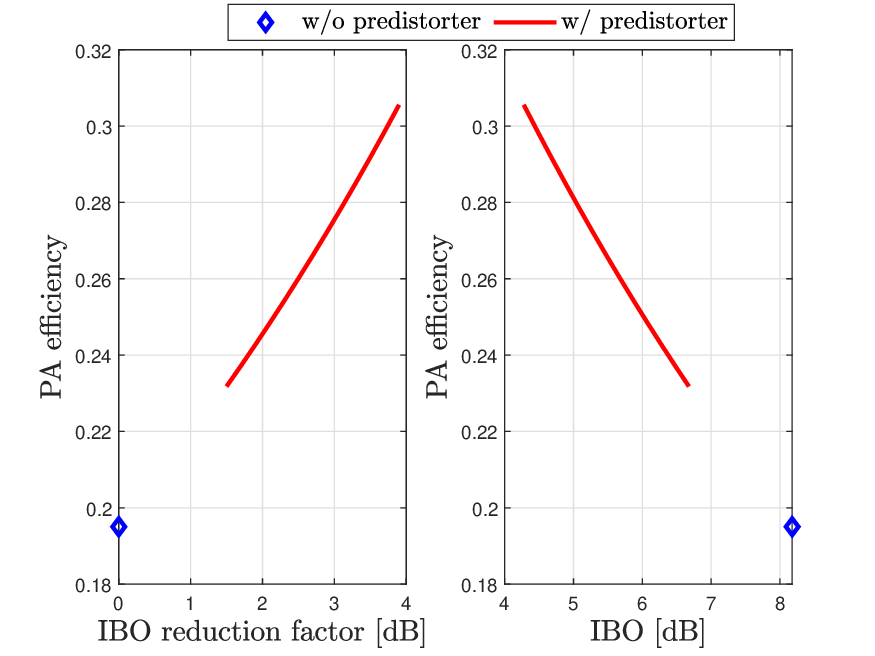}\label{fig:Predistorter_PAEfficiency}} \\
  \subfloat[EVM.] {\includegraphics[width=1\columnwidth]{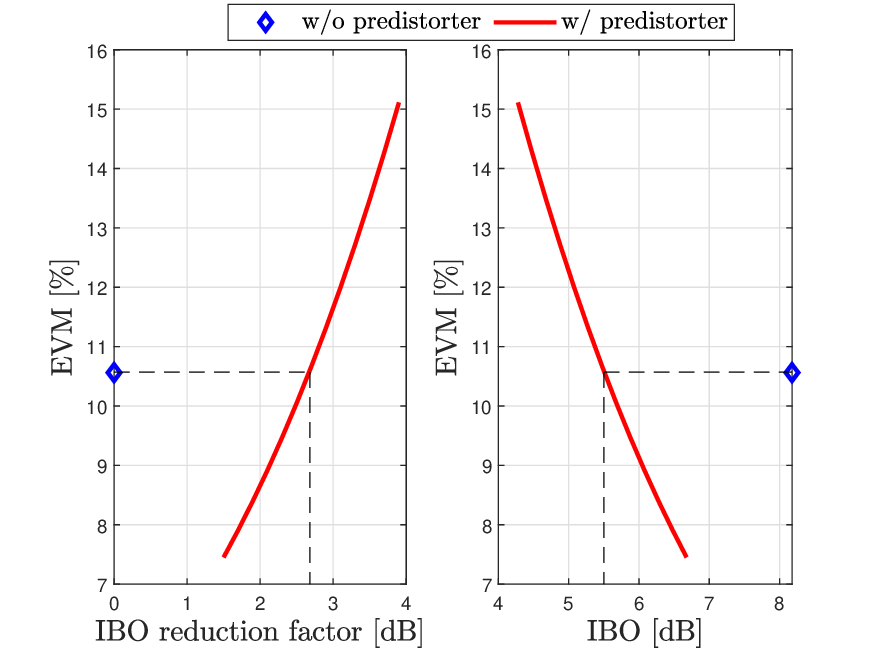}\label{fig:Predistorter_EVM}} \caption{\small PA efficiency and EVM as a function of the IBO reduction factor and IBO. } \label{Fig:PA_Eff_EVM}
  \end{figure}

Hence, the goal is that the EVM of the output signal with the predistorter equals that of the baseline OFDM signal, i.e., that in the absence of DPD. We exemplify this algorithm by using a PA model as in (\ref{eq:PA_AM/AM}) with smoothness factor $p=1.2$, and saturation output amplitude $A_s = 1$. The PA and PD responses are approximated by polynomials of orders $P=4$ and $Q=7$, respectively, whereas the polynomial coefficients are obtained by using the LS algorithm. This is illustrated in Fig. \ref{fig:Predistorter_PAEfficiency}, where the variation of EVM as a function of IBO reduction factor and IBO is represented. From these figures it can be inferred than an appropriate value for the IBO reduction factor is $2.7$ dB, allowing for an improvement in the PA efficiency.
\subsection{$\mu$-law companding technique} \label{Sec:OptimalMu}
\begin{figure}[t!]
\centering
  \subfloat[PAPR reduction of an OFDM signal.] {\includegraphics[width=1\columnwidth]{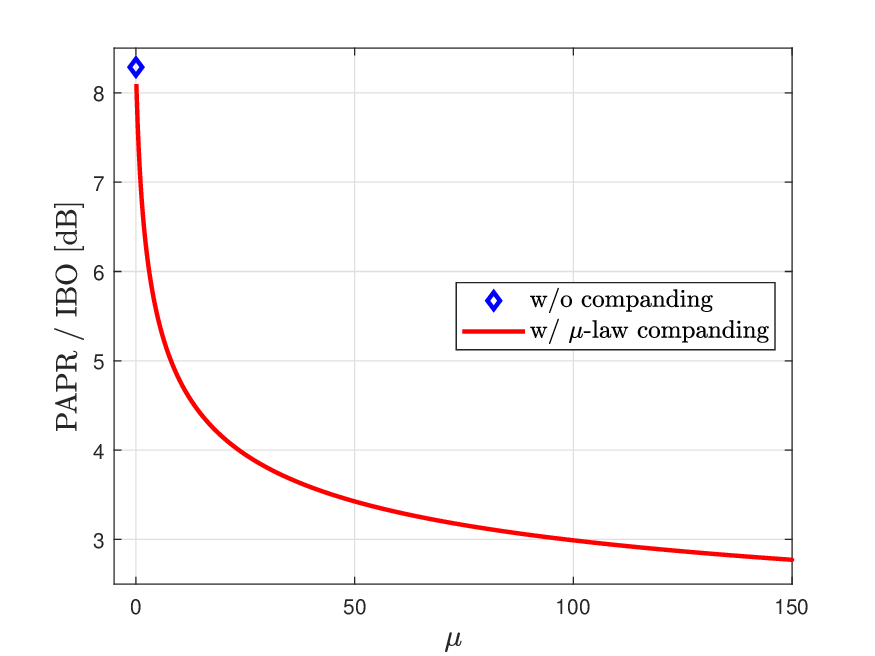}\label{fig:PAPR_muVar}} \\
  \subfloat[PA efficiency as a function of $\mu$ and IBO.] {\includegraphics[width=1\columnwidth]{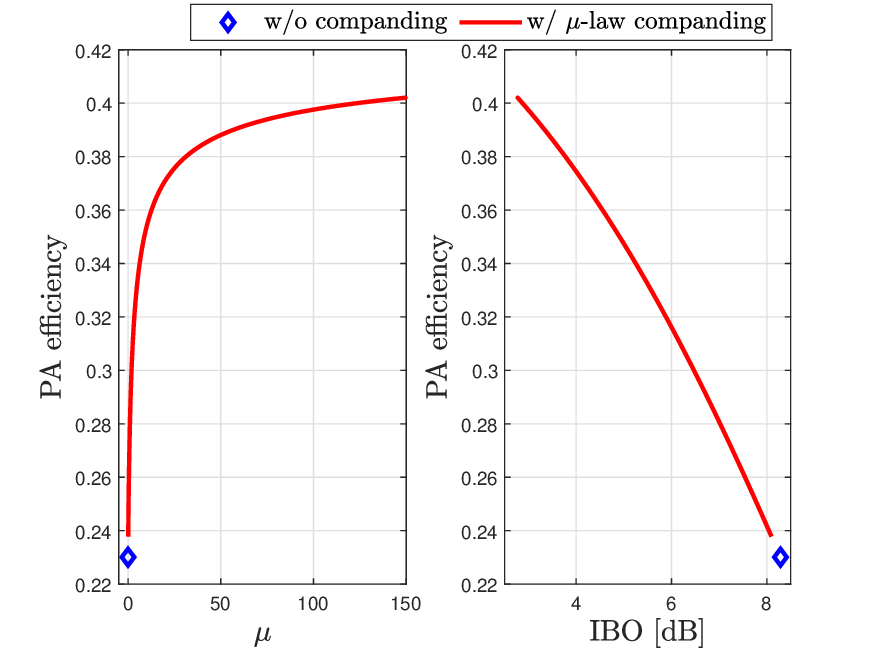}\label{fig:PA_Eff_mu_IBO}} \caption{\small PAPR level reduction and improvement of the PA efficiency using  $\mu$ law companding. } \label{Fig:muLaw_companding}
\end{figure}

The effect of a compander is a compression in the amplitude of the companded signal. Hence, as the $\mu$ parameter becomes larger, a greater compression of the input signal is achieved and the PAPR of the signal is reduced. This is exemplified in Figs. \ref{fig:PAPR_muVar} and \ref{fig:PA_Eff_mu_IBO}, where higher levels of PA efficiency are obtained as the value of $\mu$ parameter increases thanks to the companding process, while also reducing the IBO levels. We also see that the largest improvement in PA efficiency is achieved for lower values of $\mu$, while such an improvement tends to saturate as $\mu$ is further increased.

\textcolor{black}{For a sufficiently large value of $\mu$; e.g., $\mu=255$ as an analogy to speech signal applications \cite{Red_PAPR_muLaw_Wang_Tjhung}, the OFDM signal after compression tends to exhibit a PAPR similar to that of a CW signal}; this rises the efficiency of the PA and the {\color{black}EHn}, but with the consequence of losing the ability to bear information signals.  
The use of companding comes in hand with the addition of distortion due to the high non-linearity of the companding operation. Hence, a higher $\mu$ may reduce PA distortion but dramatically increases distortion due to compression introduced into the dynamic range of the signal. Therefore, the key aspect in the companding design is to find a value of $\mu$ that achieves the lowest possible overall distortion. In the following derivation, the relationship between the signal-to-noise (SNR) ratio and the distortion introduced by the companding process \textcolor{black}{is characterized}.

\begin{figure}[t!]
\centering 
\captionsetup{justification=centering}
\includegraphics[width=1\columnwidth]{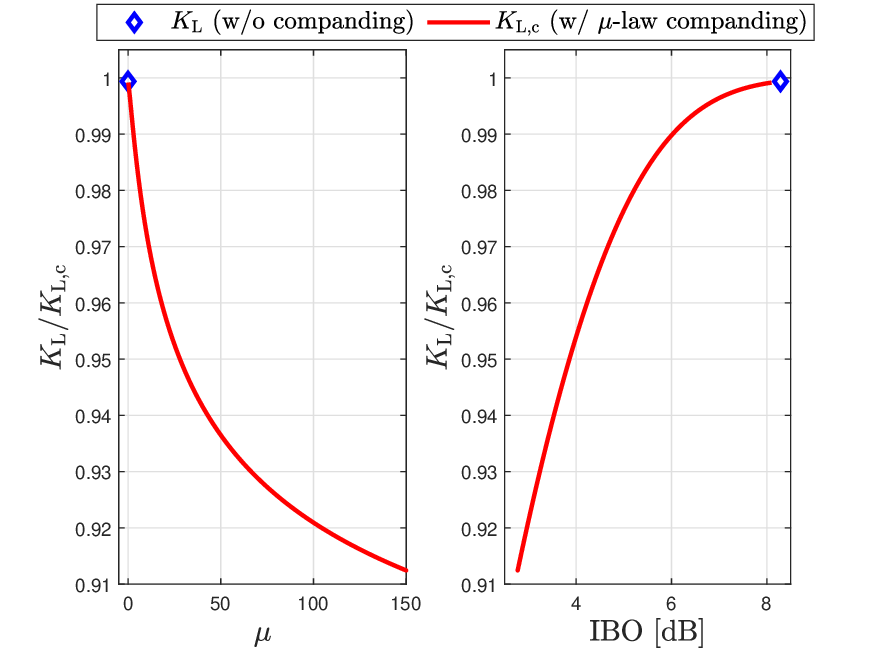}
\caption{\small Variation of $K_{\rm L,c}$ as a function of $\mu$ and IBO.}
\label{fig:KL_mu_IBO}
\end{figure}

The compressed signal obtained after the companding process, can be expressed by
\begin{equation}
    x_{\rm c}(n) = \frac{A  \ln \left(1 + \frac{\mu |x(n)|}{A}  \right)}{\ln (1 + \mu)} {\textrm{sgn}}(x(n)).
    \label{Eq:Companded}
\end{equation}
The compressed signal is then converted to time domain and amplified by the PA. This latter process introduces an additive distortion term $w_{\rm d}(t)$ with zero mean and variance $\sigma_{\rm d}^2$. The resulting signal can be expressed by
\begin{equation}
    s_{\rm c}(t) = K_{\rm L,c}  x_{\rm c}(t) + w_{\rm d}(t),
\end{equation}
\noindent where $K_{\rm L,c}$  and $\sigma_{\rm d}^2$ are obtained from Bussgang theorem as described in Section II, now using the compander transformation function instead of the PA response. The variation of $K_{\rm L,c}$ as a function of $\mu$ and IBO is illustrated in Fig. \ref{fig:KL_mu_IBO} for exemplary purposes.
At the destination node, the received signal is affected by thermal noise, inducing the addition of an AWGN term $w_{\rm a}(t)$ with zero mean and variance $\sigma_{\rm a}^2$. The received signal is then converted to the digital domain by an analog-to-digital converter (ADC), resulting in
\begin{equation}
    r(n) = K_{\rm L,c}  x_{\rm c}(n) + w_{\rm d}(n) + w_{\rm a}(n),
\end{equation}
where we assumed for simplicity an AWGN channel; this will be later generalized to include fading. After applying the expanding process, in order to recover the transmitted signal, the expanded signal $x'(n)$ can be obtained by
\begin{equation}
   x'(n) = \frac{A \exp{\left(\frac{K_{\rm L,c}  x_{\rm c}(n) + w_{\rm d}(n) + w_{\rm a}(n)}{A  \textrm{sgn}(x_{\rm c}(n))} \ln(1+\mu) \right)} - A }{\mu \ \textrm{sgn}(x_{\rm c}(n))}.
    \label{Eq:Decompanded}
\end{equation}
\noindent Now, noting that $\textrm{sgn}(x_c(n)) = \textrm{sgn}(x(n))$, and substituting (\ref{Eq:Companded}) into (\ref{Eq:Decompanded}), it is obtained
\begin{equation}
    x'(n) =  \frac{A   \left(1 + \frac{\mu |x(n)|}{A}  \right)^{K_{\rm L,c}}  \exp{ \left(\left( w_{\rm d}(n) + w_{\rm a}(n)\right)B  \right)} - A}{\mu  \textrm{sgn}(x(n))},
    \label{Eq:Decompanded_S}
\end{equation}
\noindent where $B={\ln(1+\mu)} / {A  \textrm{sgn}(x(n))}$. 
\noindent Now, using the series expansion for the exponential function, we can write
\begin{align}
    \exp{\left( \left( w_{\rm d}(n) + w_{\rm a}(n)\right) B \right)} \approx & \Biggr[ 1 + \left( w_{\rm d}(n) + w_{\rm a}(n)\right) B \nonumber \\ & \hspace{-1.1cm} \left. + \frac{\left( w_{\rm d}(n) + w_{\rm a}(n)\right)^2 B^2}{2!} + \cdots \right].
\end{align}
In cases where moderate values of IBO are used, the distortion term is usually small and the higher order terms in the above equation are much smaller than the remaining terms, which can be neglected. On the other hand, as observed in Fig. \ref{fig:KL_mu_IBO}, the value of $K_{\rm L,c}$ remains close to 1 especially when the IBO is increased, i.e., when low values of $\mu$ are used. Under these circumstances, the expanded signal can be approximated as
\begin{align}
  x'(n)  =& \frac{ \left( A + \mu |x(n)| \right)    \left(  1 + \left( w_{\rm d}(n) + w_{\rm a}(n)\right) B \right)   - A }{\mu  \textrm{sgn}(x(n))} \nonumber \\
  =& \Biggr[ x(n) + \frac{ \left( w_{\rm d}(n) + w_{\rm a}(n)\right) A B }{\mu}  \nonumber \\ 
  & \quad + x(n) \left( w_{\rm d}(n) + w_{\rm a}(n)\right) B \Biggr],    
  \label{Eq:Decompanded_Taylor}
\end{align}
\noindent where we used ${|x|} / {\textrm{sgn}(x)} = x$.

In order to recover the original data for information decoding, the expanded samples $x'(n)$ are processed by the FFT block. The equivalent frequency-domain baseband signal at subcarrier $k$ is obtained by applying the discrete Fourier transform (DFT), as expressed by

\begin{align}
    D_{\rm m}(k) =& \frac{1}{\sqrt{N}}\sum_{n=0}^{N-1} x'(n) e^{-\frac{j2\pi kn}{N}} \nonumber \\
    =& \left[\underbrace{ \frac{1}{\sqrt{N}}\sum_{n=0}^{N-1}  x(n) e^{-\frac{j2\pi kn}{N}}}_{Y(k)} \right. \nonumber \\ & \left. + \underbrace{\frac{1}{\sqrt{N}}\sum_{n=0}^{N-1} w_{\rm d}(n) \left( \frac{  A B }{\mu} + x(n) B\right) e^{-\frac{j2\pi kn}{N}}}_{D(k)} \right. \nonumber \\
    & \left. + \underbrace{\frac{1}{\sqrt{N}}\sum_{n=0}^{N-1} w_{\rm a}(n) \left( \frac{A B }{\mu} + x(n) B \right)  e^{-\frac{j2\pi kn}{N}}}_{W(k)} \right].
    \label{Eq:DFT}
\end{align}

Assuming for simplicity, yet without loss of generality, that the transmitted OFDM signal $x[n]$ is normalized (i.e., $\sigma^2_{\rm OFDM} = \mathbb{E}\left\{ |x(n)|^2 \right\} = 1$), we have $\sigma^2_{Y} = \mathbb{E}\left\{ |Y(k)|^2 \right\} = 1/N$. 

From (\ref{Eq:DFT}), the average power for the $W(k)$ term associated with the antenna noise and the average power for the $D(k)$ term associated with the distortion introduced by the PA can be computed, respectively, as
\begin{align}
     \sigma_{\rm W}^2 = & \mathbb{E}\left\{|W(k)|^2 \right\}  \nonumber \\ = & \sigma_{\rm n}^2 \left(   \left(\frac{  \ln(1+\mu) }{\mu }\right)^2  \right. +\left.  \left(\frac{  \ln(1+\mu) }{A}\right)^2       \right),
   \label{Eq:NoisePower} \\ 
   \sigma_{\rm D}^2 = & \mathbb{E}\left\{|D(k)|^2 \right\}  \nonumber \\ = &  \sigma_{\rm d,c}^2 \left(   \left(\frac{  \ln(1+\mu) }{\mu }\right)^2  \right. +\left.  \left(\frac{  \ln(1+\mu) }{A}\right)^2       \right),
   \label{Eq:DistPower}
 \end{align}
\noindent where $\sigma_{\rm d,c}^2$ is the variance of the distortion obtained from Eq. \eqref{eq:sigmad} applying the companding technique. 
 {\color{black} From Eqs. \eqref{Eq:NoisePower} and 
\eqref{Eq:DistPower}, $\mu$ (and consequently, the IBO) can be designed so as to minimize the noise and distortion effects.}
 
Finally, the SNR\footnote{For notational simplicity, distortion \textcolor{black}{is included} as an additional noise term in the SNR definition.} can be expressed by
 \begin{align}
     {\rm SNR} = & \frac{\sigma_{\rm Y}^2}{\sigma_{\rm W}^2 + \sigma_{\rm D}^2} \nonumber \\ 
     = & \frac{1 } {N \left(\sigma_{\rm a}^2 + \sigma_{\rm d,c}^2\right) \left( \left( \tfrac{ \ln(1+\mu) }{\mu} \right)^2 + \left( \tfrac{  \ln(1+\mu) }{A} \right)^2 \right) }.
 \end{align}
 
\begin{figure}[t!]
\centering 
\captionsetup{justification=centering}
\includegraphics[width=\columnwidth]{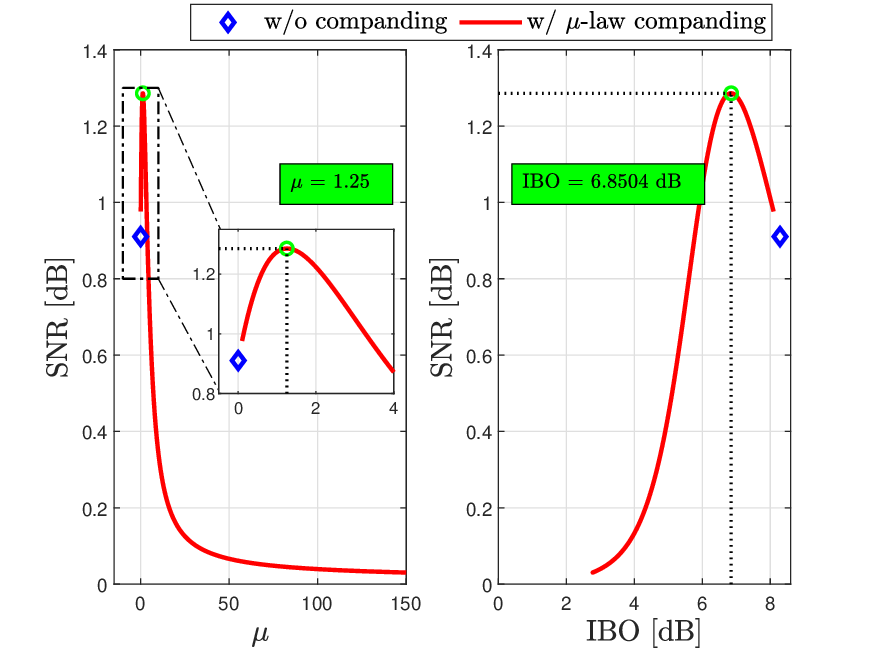}
\caption{\small SNR as a function of $\mu$.}
\label{fig:SNR_mu_IBO}
\end{figure}

 Computing the derivative of the SNR with respect to $\mu$ and equating to zero, the value of $\mu$ that maximizes the SNR can be numerically evaluated. For exemplary purposes, the variation of the SNR as a function of $\mu$ and IBO is shown in Fig. \ref{fig:SNR_mu_IBO}. 
 We see that, for a symbol size $N=512$ and the $\mu$ law compander used in the design, the SNR is maximized is approximately for $\mu \cong 1.25$.

Taking into account these results, it is possible to rewrite the equations (\ref{Eq:C_PS}) and (\ref{Eq:E_PS}) using now the expressions (\ref{eq:KL_IBO}), (\ref{eq:sigmad_IBO}), (\ref{Eq:NoisePower}) and (\ref{Eq:DistPower}). The achievable rate when the companding technique is applied can be expressed by
{\color{black}
\begin{equation}
C^{\rm C} = \log_2 \left( 1 + \frac{|h_{\rm SD}|^2 \Gamma_{\rm I} K_{\rm L,c}^2}{|h_{\rm SD}|^2 \Gamma_{\rm I} \sigma_{\rm D}^2 + (1-\rho) \sigma_{\rm W}^2 + \sigma_{\rm p}^2} \right),
\label{Eq:C_compand}
\end{equation}
}
\noindent and the harvested energy can be described by
{\color{black}
\begin{equation}
H^{\rm C}_{\rm E} = \eta_{3} \left(|h_{\rm SD}|^2 \Gamma_{\rm E} (K_{\rm L,c}^2 + \sigma_{\rm D}^2) + \rho  \sigma_{\rm W}^2 + \sigma_p^2 \right) T,
\label{Eq:E_compand}
\end{equation}
\noindent noting that the superindex $(\cdot)^C$ is aggregated to indicate that the companding technique is applied.
}

The {\color{black} PS} ratio needs to be designed so that {both} the {\color{black}EHn} and the information receiver operate over their sensitivity level; i.e., a minimum amount of power is required at {both} receivers for proper operation. We must note that the state-of-the-art RF {\color{black}EHn} sensitivity is close to $-35$ dBm \cite{Sensitive_RF_EH}. On the other hand, information receiver sensitivity ranges from $-140$ dBm (e.g., for low-bandwidth radios such as LoRa \cite{talla2017lora}) to $-85$ dBm (e.g., higher bandwidth GSM cellphones). Hence, there is a remarkable operational gap over $50$ dB between {\color{black}EHn} and information receivers \cite{NLEH_WIPT_Alevizos}. 
With this in mind, it is evident that the power splitter needs a large $\rho$ that can derive most of the signal power to the rectifier devices. In this line, a possible choice of $\rho \cong 60$ dB seems to be a reasonable estimate; this implies that only one part per million of the received signal power is diverted to the information receiving circuits\footnote{{\color{black}It is worth to note that, since the solution of a joint optimization of the {\color{black} PS} ratio $\rho$ and the companding factor $\mu$ seems very complex, we conducted an alternative approach to achieve a sub-optimal yet engineeringly-feasible solution. In this way, in the first place, we derive the optimal $\mu$ that maximizes the SNR at the information receiver. Then, for such a value, the {\color{black} PS} factor $\rho$ is designed to account for the dissimilar sensitivities of the information and EH receivers.}}.
\begin{figure}[t!]
\centering
  \subfloat[Rate $C^{\rm C}$ as a function of $\rho$.] {\includegraphics[width=1\columnwidth]{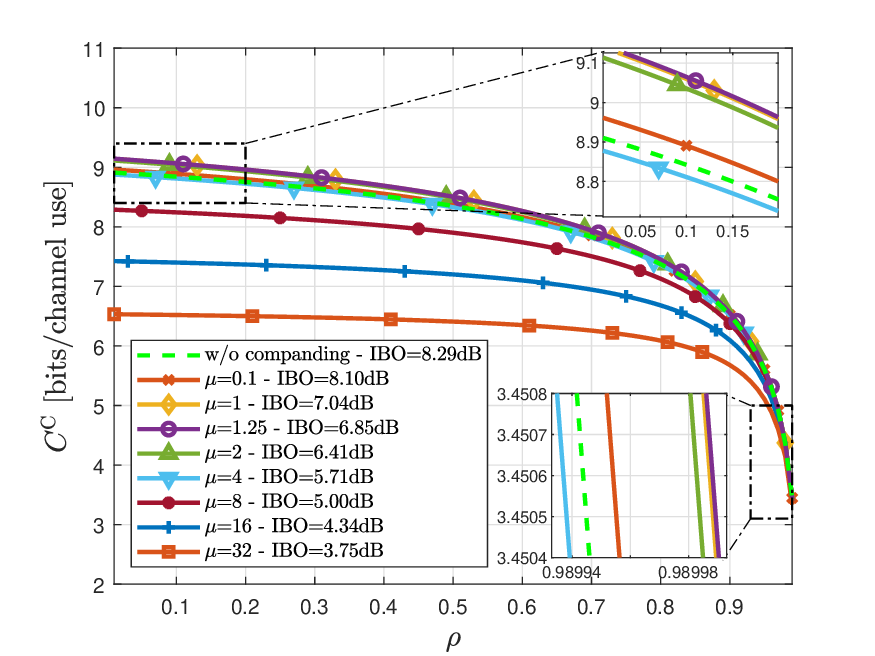}\label{fig:Rate_rho_companding_new}} \\
  \subfloat[Power/energy harvested $H^{\rm C}_{\rm P/E}$ as a function of $\rho$.] {\includegraphics[width=1\columnwidth]{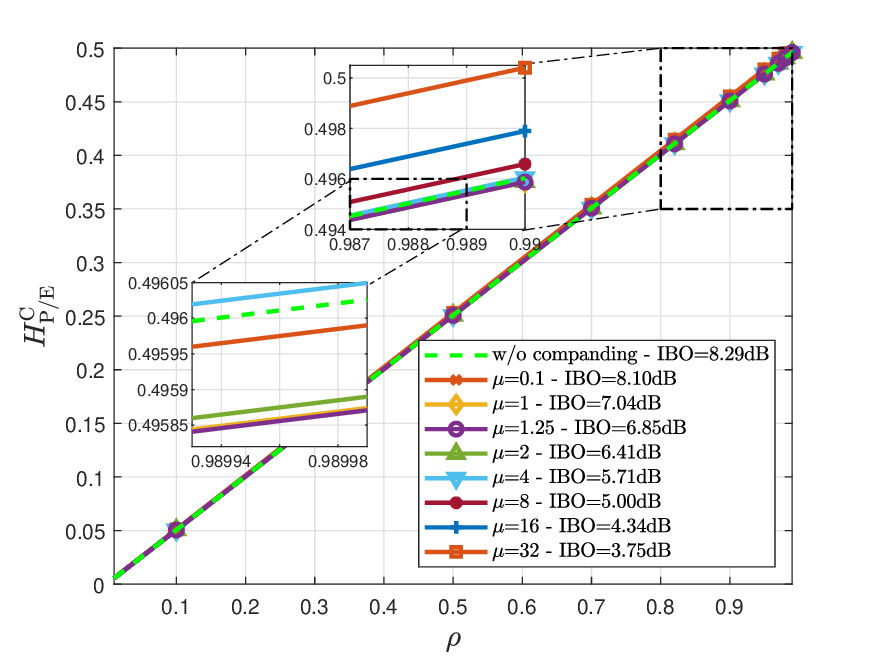}\label{fig:Q_rho_companding_new}} \caption{\small Rate and power/energy harvested functions when the companding technique is applied. With $P_{\rm{RF,Tx}} = 1$, $h_{\rm SD} = 1$, $\eta_{3,{\rm L}} = 0.5$ and $\sigma_{\rm a}^2 = \sigma_{\rm p}^2 = 10^{-3}$. } \label{Fig:Rate_Energy_Rho_NEW}
\end{figure}

The dependence of the achievable rate and the harvested energy with $\rho$ are graphically exemplified in Figs. \ref{fig:Rate_rho_companding_new} and \ref{fig:Q_rho_companding_new}. Because of the extreme value of $\rho$ required to compensate the huge difference between the {\color{black}EHn} and information receiver sensitivities, it is necessary to focus in the regions where the maximum values of $\rho$ are found. We observe  that for values of $\mu$ between $0$ and $1.25$ (the optimal compander design), the rate increases until it reaches a maximum value. Beyond this value of $\mu \cong 1.25$, the SNR starts to decrease making the rate also decrease.  
According to (\ref{Eq:E_compand}), the harvested energy rises, although marginally, with $\mu$. This is in coherence with the fact that, according to Eq. (\ref{Eq:E_compand}), all terms contribute to the sum of the energy. When the value of $\mu$ exceeds the optimal value that maximizes the SNR, the power of the term associated with the antenna noise $\sigma_{\rm W}^2$ starts to increase causing the overall power to increase. Interestingly, power levels are maintained as the IBO is reduced, which is the stark contrast with the results in Fig. \ref{fig:Power_rho_PS}, i.e., when no compansion is applied. 
\section{End-to-end performance evaluation}
\label{Sec:Overall_Eval}
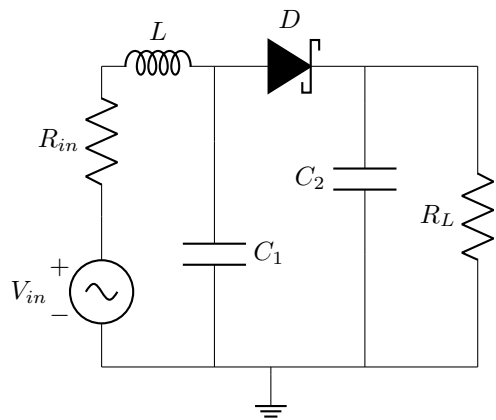
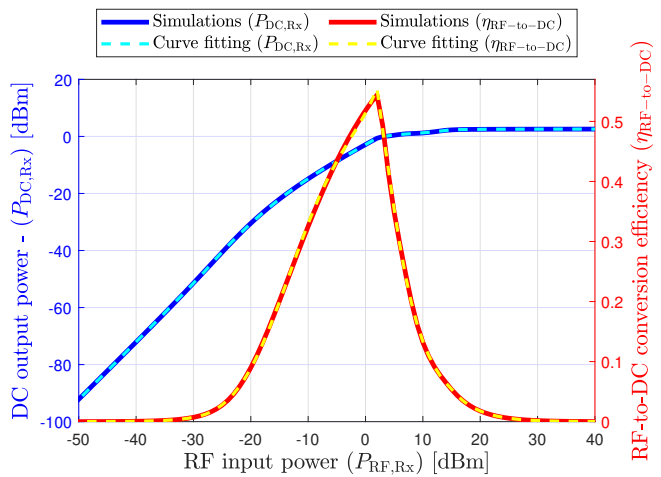
\begin{figure}[t!]
\centering
\subfloat[][Rectenna model. $R_{in}=50$ $\Omega$,  \\ $L=22.4$ nH, $C_1=1$ pF, $C_2 =2$ $\mu$F.] 
{
\label{fig:1_diodo_modelo_CircuitiKZ}
\begin{circuitikz}[american]
\draw (0,-1.5) to [vsourcesin=$V_{in}$, v<=$V_{in}$] (0,0.5); 
\draw (0,0.5) to [R=$R_{in}$] (0,2.5);
\draw (0,2.5) to [cute inductor, l=$L$] (1.5,2.5);
\draw (1.5,2.5) to [short] (1.5,1.5);
\draw (1.5,1.5) to [capacitor, l=$C_1$] (1.5,-1.5);
\draw (1.5,2.5) to [full Schottky diode, l=$D$] (3.5,2.5);
\draw (3.5,2.5) to [short] (3.5,2);
\draw (3.5,2) to [capacitor, l_=$C_2$] (3.5,0);
\draw (3.5,0) to [short] (3.5,-1.5);
\draw (3.5,2.5) to [short] (5,2.5);
\draw (5,2.5) to [R, l_=$R_L$] (5,-1.5);
\draw (5,-1.5) to [short] (0,-1.5);
\draw (2.25,-1.5) to [short](2.25,-1.6) node[ground]{};
\end{circuitikz}
} 
\\
\subfloat[][Simulation and curve fitting of RF-to-DC conversion efficiency and DC harvested power of the rectifier.] {\includegraphics[width=1\columnwidth]{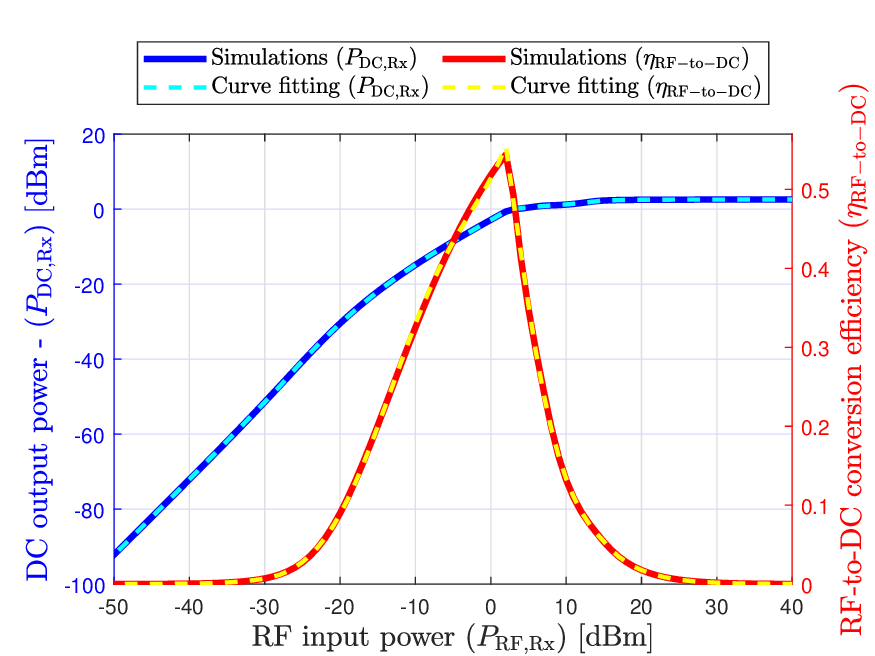}\label{fig:CurveFitting_both}}
\caption{\small Rectenna circuit model, RF-to-DC conversion efficiency and DC harvested power.}
\label{Fig:ADS_modelling}
\end{figure}

In this section, the power transfer efficiency and error performance of the SWIPT system implementing companding and DPD \textcolor{black}{is evaluated}. We take into consideration the PA and EHn efficiencies under realistic modeling, {\color{black}since the benefits of using companding and DPD are due to the non-linear behavior of the PA and the EHn.} Let us begin describing the structure for the {\color{black}EHn} under consideration. We use a single-diode rectifier circuit with the same structure as the one used in \cite{Clerckx_OnTheBeneficial}, but optimized for a carrier frequency of 915 MHz. The rectifier circuit, as represented in Fig. \ref{fig:1_diodo_modelo_CircuitiKZ}, is composed of an impedance matching circuit, a rectifying diode, and a low-pass filter. The rectifying diode is a Schottky diode SMS7630 with a low biasing voltage level, which is appropriate for rectification processes with low power levels at the input. The parameters of the rectenna are obtained by a power sweep of a single-tone input signal, i.e., a CW signal, with a frequency of $915$ MHz. The impedance matching and the low-pass filter are then designed for this frequency. In order to reach the maximum possible RF-to-DC conversion efficiency, the value for the load impedance $R_L$ is chosen as $350$ $\Omega$. Then, the values of $L$ and $C_1$ \textcolor{black}{for the} impedance matching network and the value of the output capacitor $C_2$ are calculated using the Smith Chart Utility from the Advanced Design System (ADS) software. 

In Fig. \ref{fig:CurveFitting_both}, the circuit evaluations using the ADS software for the RF-to-DC conversion efficiency and the DC harvested power are shown, as a function of the input power. Then, a curve fitting-based non-linear rectifier model is built using piecewise low-order polynomials, which is superimposed over the simulations in Fig. \ref{fig:CurveFitting_both}. This model will be later used in the simulations, to properly capture all non-linear behaviors associated to the {\color{black}EHn}.

Without loss of generality, we assume that the distance between the source and the destination is $1$m, which is in the far-field region for the frequency of operation. Hence, the path loss experienced by the signal equals ${\rm PL}_0 = 20 \log(\lambda/4\pi)$, where $\lambda$ is the signal wavelength. The OFDM signal is composed by $N=512$ subcarriers with subcarrier spacing {$\Delta_{\rm f}=15$ kHz}, and the modulation scheme per subcarrier is quadrature-phase-shift-keying (QPSK). The following set of experiments is carried out over different types of channels, i.e. AWGN, Rice, single-tap Rayleigh, and multi-tap Rayleigh channels. Throughout all experiments, the PS factor $\rho=60$ dB is used, to compensate for the different sensitivity levels of the information and {\color{black}EHn} receivers.
\subsection{AWGN channel}
The system performance assuming an AWGN channel \textcolor{black}{is now analyzed}, which serves as an upper bound in terms of performance. First, we consider the cases on which DPD and companding are used alone to better understand their effects on PA and {\color{black}EHn} efficiencies. Then, the combination of both techniques is evaluated.
\subsubsection{DPD technique}
\begin{figure}[t!]
\centering {\includegraphics[width=1\columnwidth]{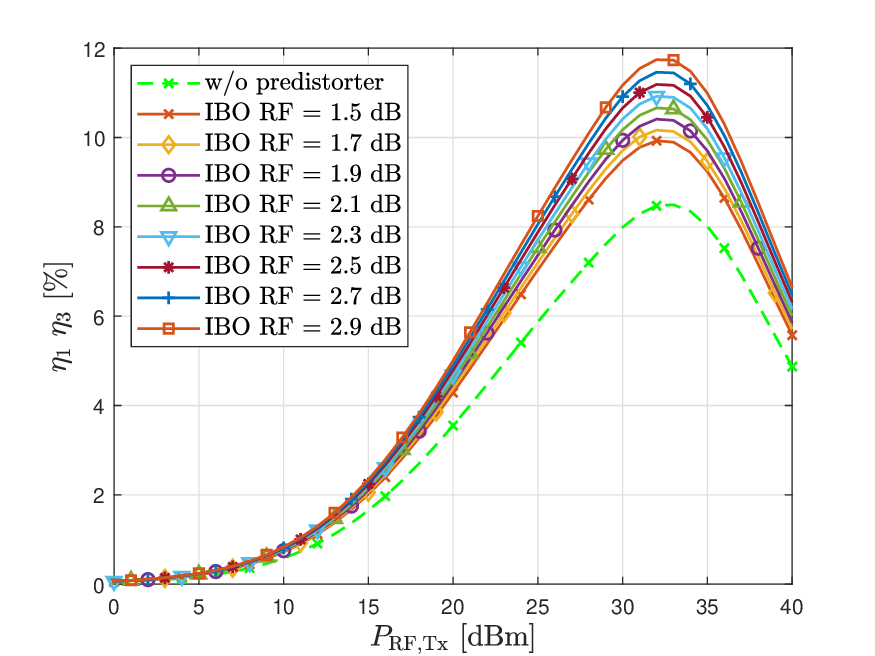}}
\caption{\small End-to-end {\color{black}power transfer} efficiency for the DPD technique.}
\label{fig:Predistorter_EtE_EEff_eta1_eta3}
\end{figure}

\begin{figure}[t!]
\centering
\subfloat[ {\color{black}EHn} efficiency for different IBO reduction factors.] {\includegraphics[width=1\columnwidth]{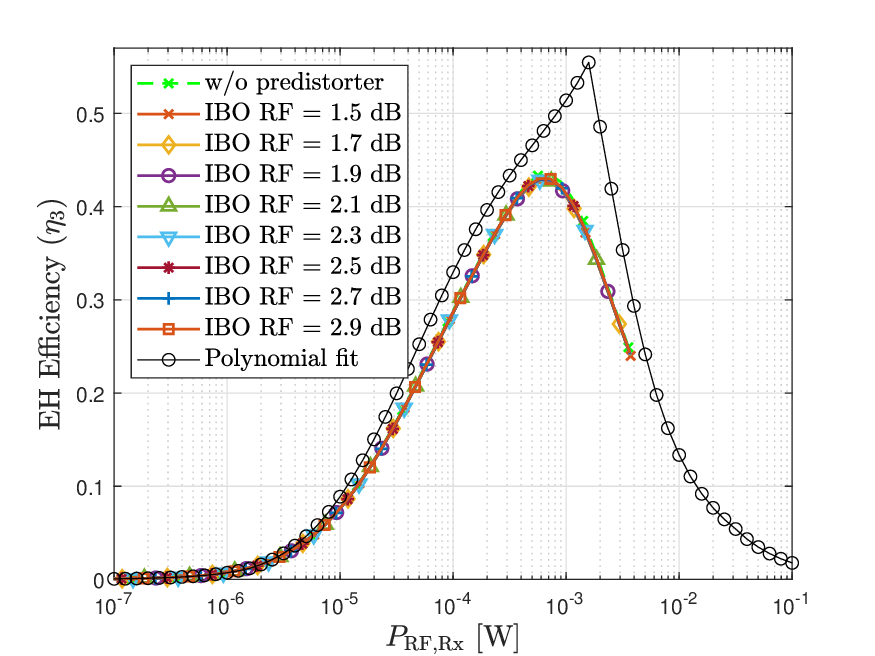}\label{fig:Predistorter_EHEff_new}} \\
\subfloat[BER for different IBO reduction factors.] {\includegraphics[width=1\columnwidth]{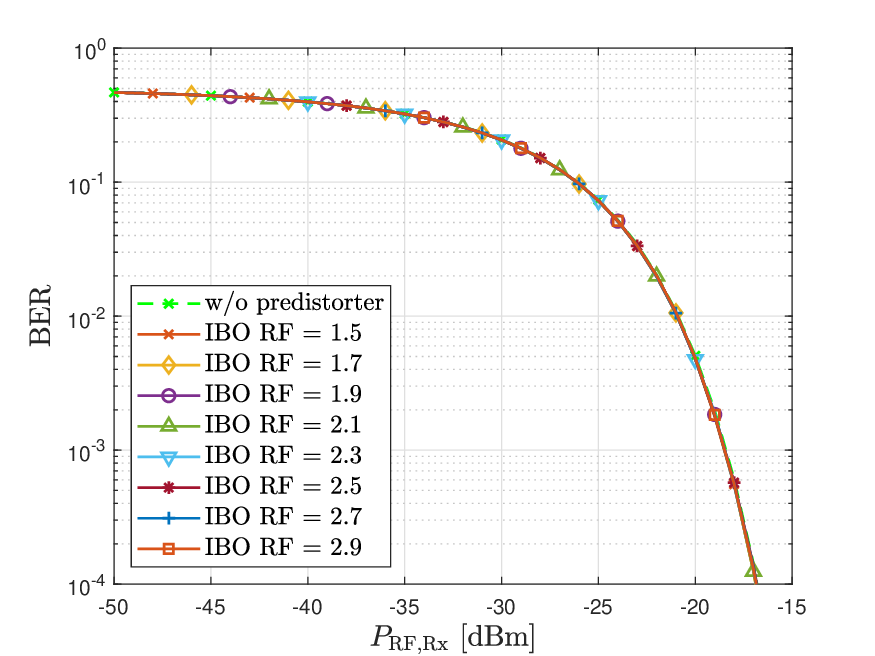}\label{fig:BER_Predistorter}} 
\caption{\small Effect of increasing the IBO reduction factor in DPD.} \label{Fig:EndToEndEfficiency_Predistorter}
\end{figure}

We consider a PA and DPD as those assumed in Section \ref{Sec:Optimal_IBO_factor}. Hence, for the set of parameters considered, an IBO reduction factor of about $2.7$ dB allows the same EVM as in the absence of DPD. By choosing this IBO, the PA efficiency is improved around a $6.8$\%. The overall end-to-end {\color{black}power transfer} efficiency\footnote{For simplicity of discussion, we dropped $\eta_2$ from the end-to-end {\color{black}power transfer} efficiency evaluation, since its effect is only dependent on the distance between source and destination.} is represented in Fig. \ref{fig:Predistorter_EtE_EEff_eta1_eta3}, as the product between PA efficiency ($\eta_1$) and the {\color{black}EHn} efficiency ($\eta_3$). We observe that the use of a DPD allows for improving the end-to-end {\color{black}power transfer} efficiency. {\color{black}We also see that as the IBO reduction factor increases, the end-to-end power transfer efficiency increases}\footnote{{\color{black}For the sake of readability, the IBO reduction factor is abbreviated as IBO
RF.}}. However, this efficiency gain is caused \textit{only} by the improvement in the PA efficiency since, as is illustrated in Fig. \ref{fig:Predistorter_EHEff_new}, the efficiency of the {\color{black}EHn} is not affected by the DPD. This is coherent with the fact that, by linearizing the PA through DPD, non-linearities of the PA are minimized. This can also be confirmed when observing the BER values in \ref{fig:BER_Predistorter}, which are not modified because of the DPD operation as the overall EVM is barely affected compared to the reference case of no DPD.
\subsubsection{Companding technique}
\begin{figure}[t!]
\centering
\subfloat[{\color{black}EHn} efficiency for different values of $\mu$.  ]{\includegraphics[width=1\columnwidth]{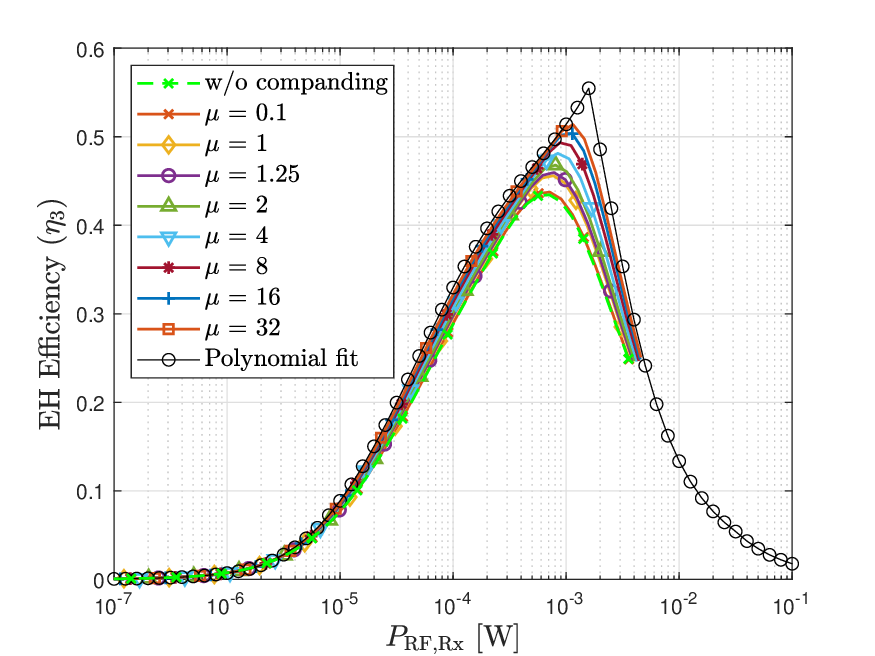}\label{fig:EH_Eff_comp}} \\
\subfloat[BER for different values of $\mu$. ]{\includegraphics[width=1\columnwidth]{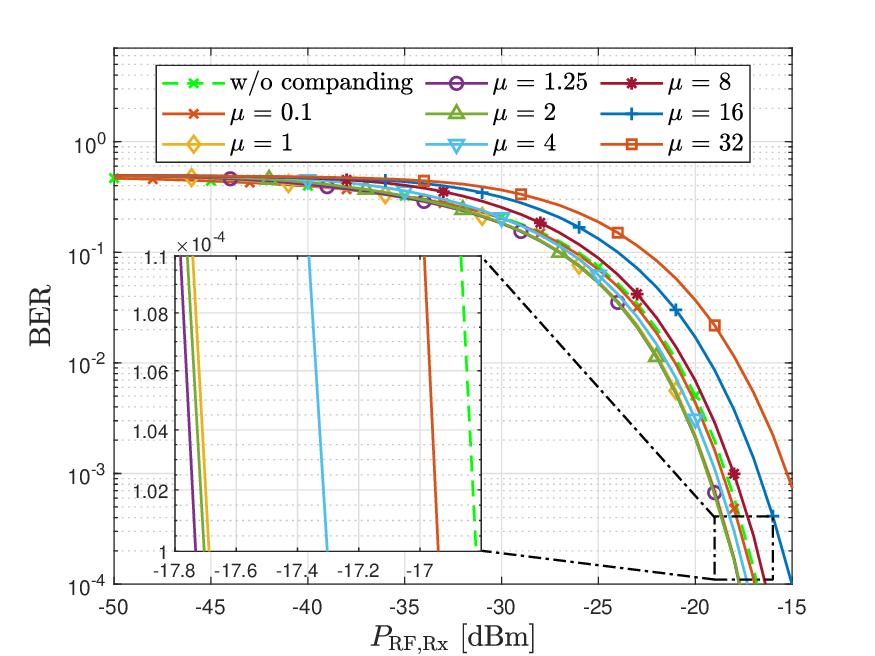}\label{fig:BER_Companding}}
\caption{\small Effect of increasing the $\mu$ companding parameter.}
\label{Fig:PAeff_harvesPower_comp}
\end{figure}

Following the same derivation as in Section \ref{Sec:OptimalMu}, the optimal value $\mu=1.25$ that maximizes the SNR \textcolor{black}{is set}. With this value, a reduction of the IBO of around $1.4$ dB \textcolor{black}{is achieved}, which translates into an increment of the PA efficiency of around $3.4$\%. Now, to evaluate the efficiency of the {\color{black}EHn} when companding is performed over the signal, we use the piecewise polynomial fitting model previously described. In Fig. \ref{fig:EH_Eff_comp}, the {\color{black}EHn} efficiency is represented for different values of $\mu$. 

At a first glance, it may be concluded that the {\color{black}EHn} efficiency is improved when the parameter $\mu$ is increased. However, this may come at the price of a degraded performance for information transfer. This is observed in Fig. \ref{fig:BER_Companding}, on which the best BER performance is obtained for the trade-off value of $\mu=1.25$. This is in coherence with the observations made in Fig. \ref{fig:Rate_rho_companding_new}. We also see that the use of companding not only improves the efficiency of the EH, but also the BER performance.

\begin{figure}[t!]
\centering {\includegraphics[width=1\columnwidth]{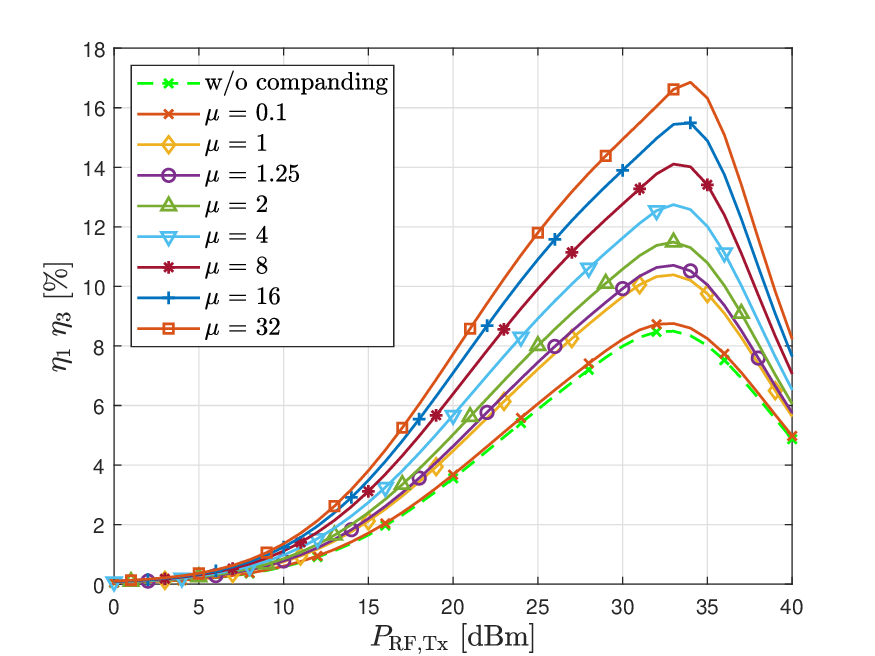}}
\caption{\small End-to-end {\color{black}power transfer} efficiency for the $\mu$-law companding technique.}
\label{fig:Companding_EtE_EEff_eta1_eta3}
\end{figure}

Finally, the end-to-end {\color{black}power transfer} efficiency due to the companding technique is represented in Fig. \ref{fig:Companding_EtE_EEff_eta1_eta3}. Again, the best efficiency is achieved for high values of $\mu$; however, the optimal trade-off value for $\mu$ yields also an increased end-to-end efficiency.
\subsubsection{Combining DPD and companding}
\label{Sec:Combination}
\begin{figure}[t!]
\centering
\subfloat[][{\color{black}EHn} efficiency for different techniques.  ]{\includegraphics[width=1\columnwidth]{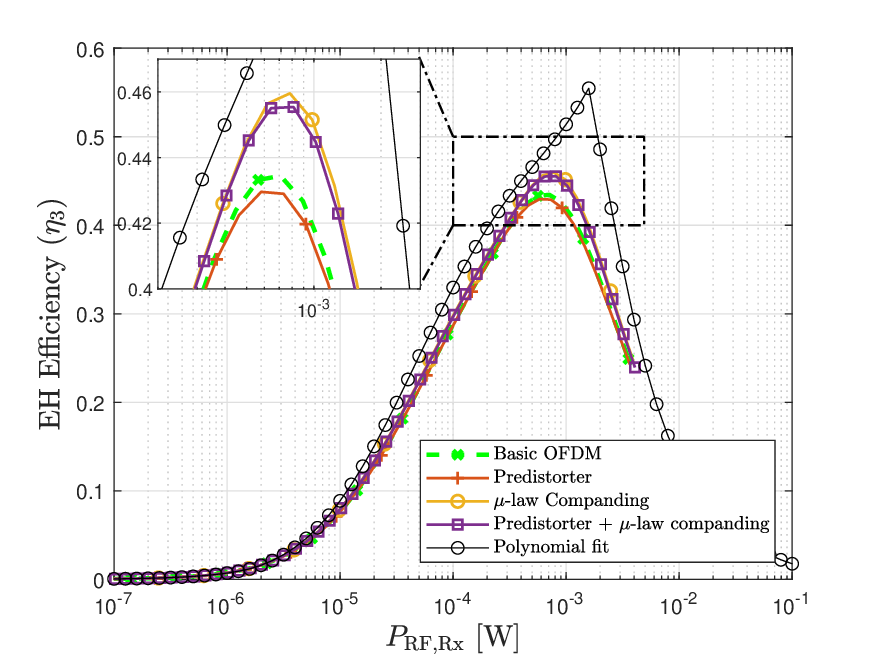}\label{fig:EH_Eff_Predistorter_companding}} \\
\subfloat[][BER for different techniques. ]{\includegraphics[width=1\columnwidth]{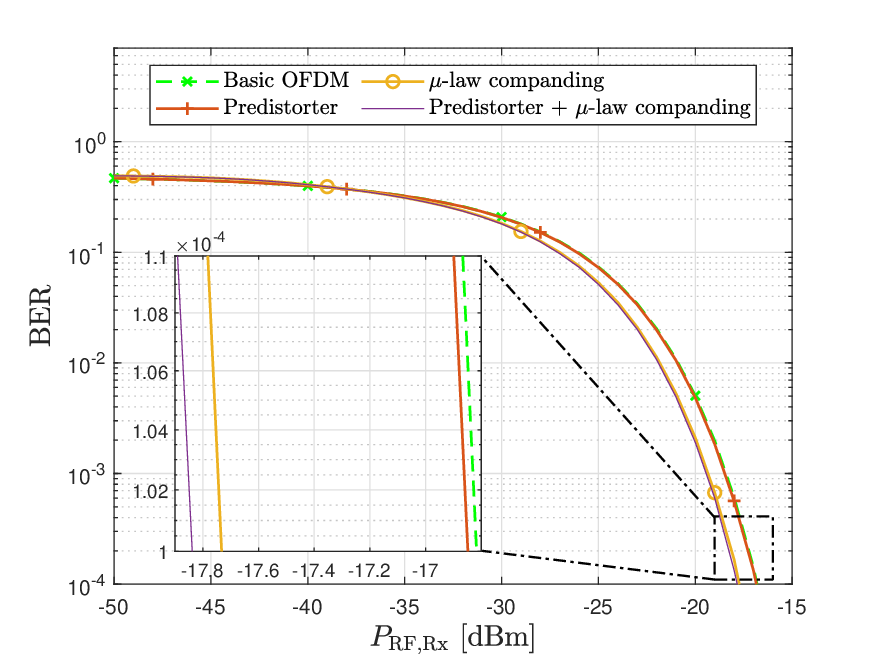}\label{fig:BER_Predistorter_Companding}}
\caption{\small Comparing EH efficiency and BER performance for DPD, companding, and joint DPD+companding.}
\label{Fig:EHeff_harvesPower_Predistorter_companding}
\end{figure}

\begin{figure}[t!]
\centering {\includegraphics[width=1\columnwidth]{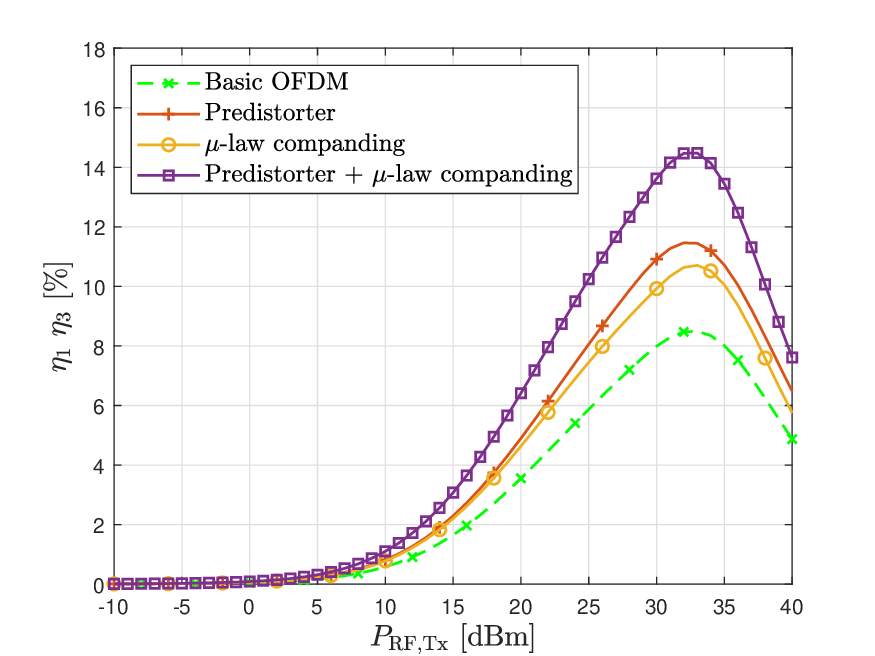}}
\caption{\small End-to-end efficiency for DPD, companding, and joint DPD+companding.}
\label{fig:Predistorter_Companding_EtE_EEff_eta1_eta3}
\end{figure}

We now evaluate the performance when both DPD and companding techniques are combined. By considering the previously derived values of $\mu=1.25$ and linearization parameters for the DPD, the combination of DPD and companding allows an IBO reduction of almost $4$ dB, which improves the PA efficiency in a $11.9$\%. While the {\color{black}EHn} efficiency is not affected by DPD, this can also be improved by using the companding technique as shown in Fig. \ref{fig:EH_Eff_Predistorter_companding}. We also see in Fig. \ref{fig:BER_Predistorter_Companding} that the BER performance is improved when both techniques are jointly used. We also observe that the end-to-end efficiency is improved when DPD and companding are combined, as confirmed by Fig. \ref{fig:Predistorter_Companding_EtE_EEff_eta1_eta3} and the values listed in Table \ref{tab:Summary}.
\begin{table*}[t!]
\tiny
\caption{\small {\color{black}Power transfer} efficiency improvements - AWGN channel.
EH efficiency measured with $P_{\rm RF,Rx} = 0$ dBm. End-to-end {\color{black}power transfer} efficiency measured with $P_{\rm RF,Tx} = 30$ dBm.}
\centering
\resizebox{0.65\textwidth}{!}{%
\begin{tabular}{|l|l|l|l|}
\hline
 &   {\color{black}PA} Eff. ($\eta_1$) [\%]&  {\color{black}EHn} Eff. ($\eta_3$) [\%]& $\eta_1$ $\eta_3$ [\%]  \\ \hline
 Baseline OFDM&  $19.9$ &  $43.4$ &  $8.6$ \\ \hline
 Predistorter&  $26.7$ &  $42.9$ &  $11.4$  \\ \hline
 $\mu$-law companding&  $23.3$ &  $45.9$ &  $10.7$ \\ \hline
 Predistorter + $\mu$-law companding&  $31.8$ &  $45.5$ & $14.5$ \\ \hline
\end{tabular}
}
\label{tab:Summary}
\end{table*}
\subsection{Fading channel}
\begin{figure}[t!]
\centering
\includegraphics[width=1\columnwidth]{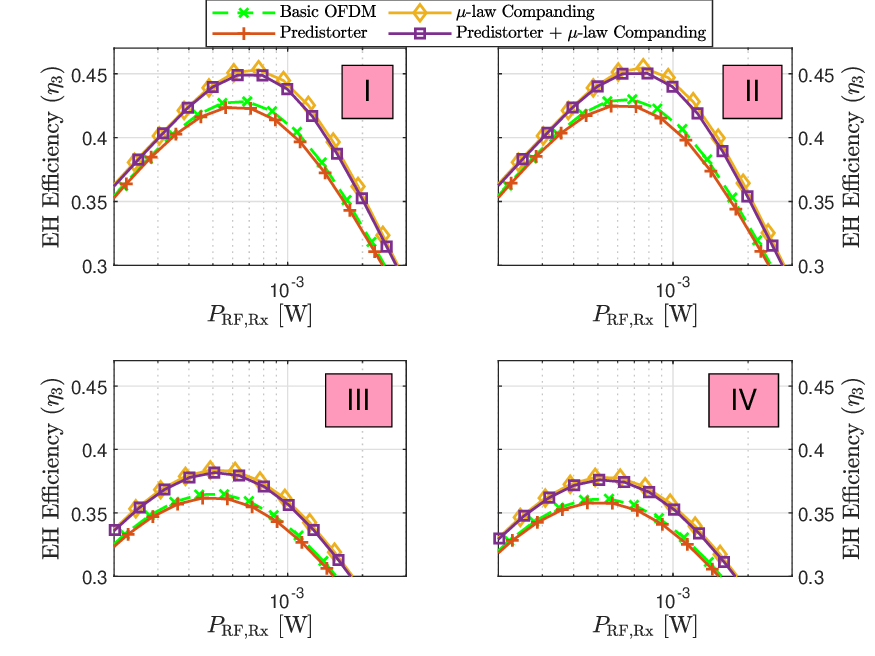}
\caption{\small {\color{black}EHn} efficiency ($\eta_3$) over fading channels.}
\label{Fig:EH_Eff_diff_chann}
\end{figure}

\begin{figure}[t!]
\centering
{\includegraphics[width=1\columnwidth]{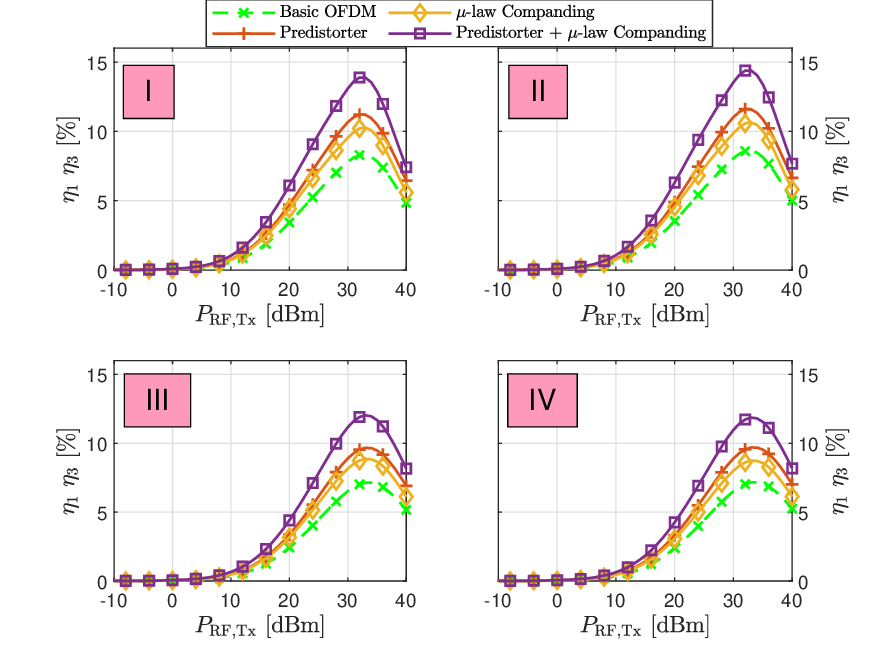}}
\caption{\small End-to-end {\color{black}power transfer} efficiency over fading channels.}
\label{Fig:eta1_eta3}
\end{figure}

\begin{figure}[t!]
\centering
{\includegraphics[width=1\columnwidth]{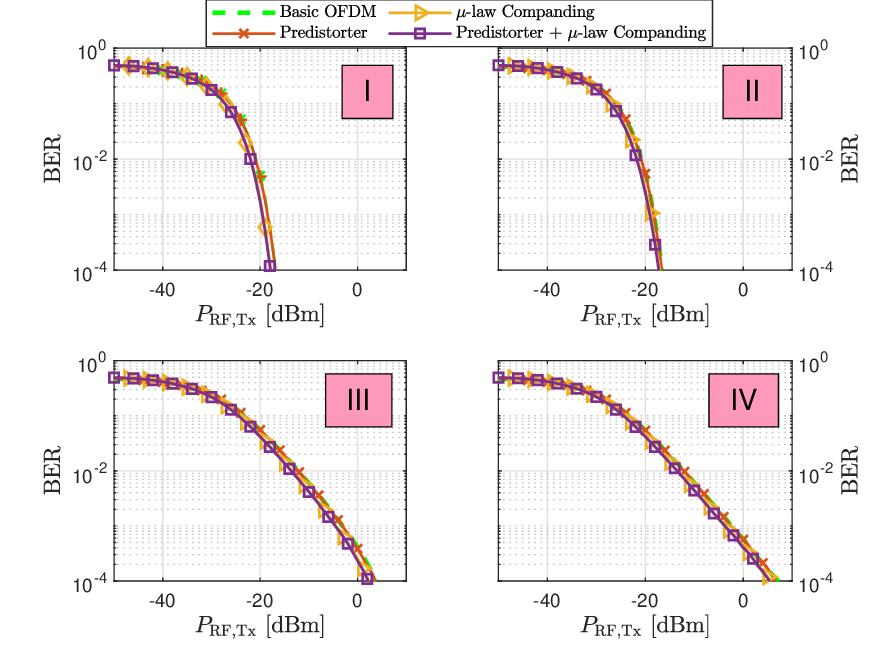}}
\caption{\small BER performance over fading channels.}
\label{Fig:BER}
\end{figure}

Even though the AWGN channel serves as a good reference for benchmarking purposes, the performance in the presence of fading also needs to be evaluated. We consider different cases of fading channels, as follows: \textit{i}) Flat Rice channel, with a strong line-of-sight (LOS) component {$K=20$ dB}, \textit{ii}) Flat Rayleigh channel, i.e, a non LOS component, and \textit{iii}) Multiple-tap frequency selective Rayleigh channel. For this latter case, a power delay profile with three taps and relative power $[0, -10, -20]$ dB is considered. In the next set of figures, {\color{black}EHn} efficiency (Fig. \ref{Fig:EH_Eff_diff_chann}), end-to-end {\color{black}power transfer} efficiency (Fig. \ref{Fig:eta1_eta3}) and error rate performance (Fig. \ref{Fig:BER}) \textcolor{black}{are evaluated} for the set of channels previously described. Colored legends indicate: \colorbox{MyPink}{I}. AWGN, \colorbox{MyPink}{II}. Rice, \colorbox{MyPink}{III}. Single-tap Rayleigh and \colorbox{MyPink}{IV}. Multi-tap Rayleigh channels.

Fig. \ref{Fig:EH_Eff_diff_chann} shows that comparable EH efficiencies are obtained for the AWGN and Rice channel cases, given the strong LOS nature of the communication. The EH efficiency is degraded in the case of Rayleigh channels, a bit more when assuming frequency-selectivity, due to the higher channel variability. We also see that the use of companding is beneficial in all instances to improve EH efficiency. We observe that the use of DPD barely affects this performance metric, regardless of whether companding is used or not. However, the end-to-end {\color{black}power transfer} efficiency is improved when {both} DPD and companding techniques are used, as observed in Fig. \ref{Fig:eta1_eta3}. This confirms the benefits of the proposed techniques to improve both the PA efficiency and the RF-to-DC conversion in the {\color{black}EHn}. Again, AWGN and Rician cases provide similar performances, while some degradation is observed in the cases of Rayleigh fading. Finally, we see in Fig. \ref{Fig:BER} little differences in terms of error performance due to DPD and companding, compared to the absence of both techniques. Indeed, frequency-selective Rayleigh case provides the worst BER performance, but this is not degraded due to the use of companding.
\textcolor{black}{\subsection{Complexity}}
\textcolor{black}{While both DPD and companding techniques seem to offer appealing features for SWIPT, \textcolor{black}{the impact of these techniques in complexity needs to be discussed}. The implementation of DPD requires that the RF signal at the output of the PA is sensed, down-converted to baseband, sampled and processed, being necessary the addition of an observation path for the estimation of the DPD parameters. This extra path implies some additional components as: RF sampler, attenuator, down-converter and ADC. This is translated into additional complexity and power consumption, which are ultimately determined by the ADC sampling rate and resolution. Furthermore, in order to better capture the non-linear effects of the PA, it is necessary to strongly oversample the signal, which implies a large resolution of the ADC and therefore high cost and power consumption. These implementation costs create a trade-off that must be taken into account to assess the convenience of using the proposed DPD technique and, even more so, when the {\color{black}power transfer} efficiency of the system is to be improved.}

\textcolor{black}{As pointed out in \cite{libro_fer_etal}, the overall power consumption of the transceiver is dominated by the PA when a large transmitted power is considered -- this is often the case in the context of SWIPT. In this situation, the implementation of DPD techniques actually allows for power saving, since the PA is able to operate with a better efficiency. Finally, it is necessary to mention that the additional complexity due to DPD is only incurred at the PB. Hence, the receive nodes do not need any additional complex circuitry, which is an important aspect given in IoT environments with energy-constrained receiver nodes.}

\textcolor{black}{On the other hand, companding techniques are attractive due to their low implementation complexity, as they can be easily implemented using simple digital signal processing blocks. Different from DPD techniques, the expansion operation requires some additional processing at the receiver nodes. However, such digital signal processing can be easily performed by the digital blocks embedded in the receiver node with a negligible cost. As a key limitation, in the case of highly frequency-selective channels an increased distortion can be observed in the received information signal after expansion \cite{companding_FS_channnels}. However, in WPT scenarios the channel between source and destination usually has a strong LOS component so that its frequency selectivity is limited. Hence, the companding technique can be used to improve WPT efficiency without degrading information transfer performance, and without incurring in a relevant increase in complexity.
}
\section{Concluding remarks}
\label{Sec:Conclusions}
The use of companding techniques to improve end-to-end {\color{black}power transfer} efficiency in SWIPT systems, and its combination with {\color{black}DPD} techniques, was proposed and analyzed for the first time in the literature. Thanks to the IBO reduction achieved by both techniques, it \textcolor{black}{was} possible to improve the PA and {\color{black}EHn} efficiencies while keeping information transfer performance unchanged. The combination of companding and DPD \textcolor{black}{was shown to bring} additional improvements, compared to when these are used separately, and \textcolor{black}{was} barely affected by fading and moderate frequency selectivity. Following the design guidelines here described, DPD and companding techniques barely impacted the information transfer performance. Hence, the use of these techniques is a good strategy to improve the efficiency of WPT when using OFDM signals in SWIPT systems.

\vspace{-3mm}
\bibliographystyle{IEEEtran}

\bibliography{References_TGCN}
\end{document}